\documentclass[prd,aps,eqsecnum,floatfix,nofootinbib,preprint,tightenlines]{revtex4}

\usepackage{latexsym}
\usepackage{graphicx}
\usepackage{multirow}
\usepackage{amsmath}

\def\floatcaption#1#2{ \caption{#2 \label{#1}} }

\def\bibi{\bibitem}

\let\inodot=\i

\def\a{\alpha}
\def\b{\beta}
\def\c{\chi}
\def\d{\delta}
\def\e{\epsilon}                
\def\f{\phi}                    
\def\g{\gamma}

\def\i{\iota}

\def\l{\lambda}
\def\m{\mu}
\def\n{\nu}
\def\o{\omega}
\def\p{\pi}                     
\def\r{\rho}                    
\def\t{\tau}

\def\x{\xi}
\def\z{\zeta}

\def\G{\Gamma}

\def\L{\Lambda}
\def\O{\Omega}
\def\P{\Pi}

\def\cd{{\cal D}}

\def\ck{{\cal K}}
\def\cl{{\cal L}}

\def\co{{\cal O}}

\def\cw{{\cal W}}

\def\cbo{{\,\raise-.15ex\Sc [\,}}                       

\def\vev#1{\Big\langle #1 \Big\rangle}           

\def\svev#1{\left\langle #1\right\rangle}       

\def\ddt#1{{\buildrel {\hbox{\LARGE .\kern-2pt.}} \over {#1}}}

\def\ie{\mbox{\it i.e.}}

\def\tr{{\rm tr}\,}

\def\half{{1\over 2}}
\def\Re{{\rm Re\,}}

\def\det{{\rm det}}

\def\ttl#1{{\it #1}}

\def\vev#1{\langle #1 \rangle}

\long\def\symbolfootnote[#1]#2{\begingroup%
\def\thefootnote{\fnsymbol{footnote}}\footnote[#1]{#2}\endgroup}

\long \def \blockcomment #1\endcomment{}

\def\seef{{\it cf.\  }}

\def\eb{s}

\def\ebb{\bar{s}}
\def\bs{\bar{s}}

\def\bc{\overline{C}}
\def\bct{\overline{c}}  

\def\bX{{\overline{X}}}

\def\bz{{\overline{\z}}}

\def\tg{\tilde{g}}
\def\tb{\tilde{\b}}
\def\tP{\tilde{P}}

\def\ts{\tilde{\cal S}}
\def\ttil{\tilde{T}}

\def\tL{\tilde{\L}}
\def\tX{{\widetilde{X}}}
\def\tZ{{\widetilde{Z}}}
\def\hO{\hat{\O}}

\def\hm{\hat\mu}

\def\aop{\tilde{A}}
\def\Zred{Z_{\rm red}}

\begin{document}

\begin{center}
\vspace{10mm}
{\large\bf
Phase with no mass gap in non-perturbatively gauge-fixed\\[3mm]
Yang--Mills theory
}
\\[12mm]
Maarten Golterman$^a$%
\symbolfootnote[2]{Permanent address: Department of Physics and Astronomy,
San Francisco State University, San Francisco, CA 94132, USA}
\ \ and \ \ Yigal Shamir$^b$
\\[8mm]
{\small\it
$^a$Institut de F\'\inodot sica d'Altes Energies, Universitat Aut\`onoma de Barcelona,\\ E-08193 Bellaterra, Barcelona, Spain}%
\\[5mm]
{\small\it $^b$School of Physics and Astronomy\\
Raymond and Beverly Sackler Faculty of Exact Sciences\\
Tel-Aviv University, Ramat~Aviv,~69978~ISRAEL}%
\\[10mm]
{ABSTRACT}
\\[2mm]
\end{center}

\begin{quotation}
An equivariantly gauge-fixed non-abelian gauge theory is a theory in which
a coset of the gauge group,
not containing the maximal abelian subgroup,
is gauge fixed.   Such theories are non-perturbatively well-defined.
In a finite volume,
the equivariant BRST symmetry guarantees that
expectation values of gauge-invariant operators are equal to their values
in the unfixed theory.  However, turning on a small breaking of this symmetry,
and turning it off after the thermodynamic limit
has been taken, can in principle reveal new phases.  In this paper we use
a combination of strong-coupling and mean-field techniques to study
an $SU(2)$ Yang--Mills theory equivariantly gauge fixed to a $U(1)$ subgroup.
We find evidence for the existence of a new phase in which two of the gluons
becomes massive while the third one stays massless,
resembling the broken phase of an $SU(2)$ theory with an adjoint Higgs field.
The difference is that here this phase occurs in an
asymptotically-free theory.
\end{quotation}

\renewcommand{\thefootnote}{\arabic{footnote}} \setcounter{footnote}{0}

\newpage
\section{\label{Intro} Introduction}
Some years ago, we proposed a new approach to discretizing non-abelian
chiral gauge theories, in order to make them accessible to the methods
of lattice gauge theory \cite{gfx}.   An essential ingredient in this new
approach is the inclusion of a gauge-fixing action at the level of
the path integral defining the theory.  Of course, whatever
gauge-fixing action one chooses, the path integral has to be well-defined
non-perturbatively.   It was shown in Ref.~\cite{HNnogo} that this is
impossible if one insists on maintaining BRST invariance:
In a fully gauge-fixed
lattice gauge theory the BRST symmetry causes the partition function, as
well as all (un-normalized) expectation values of gauge-invariant operators,
to vanish.

The problem can be circumvented if, instead of the full non-abelian group $G$,
one gauge fixes only a coset
$G/H$, where $H$ is a subgroup containing the maximal abelian subgroup
of $G$.   Building on earlier work by Schaden \cite{MS1} for the group
$SU(2)$, we showed that for $G=SU(N)$ and $H$ the maximal abelian subgroup
itself, partially (``equivariantly'') gauge-fixed Yang--Mills theories can be constructed
on the lattice satisfying an invariance theorem:
In such theories, expectation values of gauge-invariant operators
do not vanish, and are exactly equal to those in the unfixed theory, in any finite volume \cite{gfx}.  Since only the coset $G/H$ is gauge-fixed, ghosts
are only introduced for the coset generators.
Such theories are then invariant under an equivariant BRST (eBRST) symmetry,
and necessarily contain four-ghost interaction terms.   Both eBRST symmetry and the appearance of four-ghost interactions are key ingredients
for the proof of the invariance theorem.   This eBRST symmetry
is nilpotent on operators which are gauge invariant under the subgroup
$H$, guaranteeing perturbative unitarity to all orders in much the same
way that standard BRST symmetry does in the usual case \cite{gfx}.
The key difference is that, contrary to the usual gauge-fixing procedure
which is inherently perturbative \cite{HNnogo},
the equivariantly gauge-fixed theory is well defined non-perturbatively.%
\footnote{For an interesting perspective
on the possible role of $U(1)$ (sub)groups in this context, see Ref.~\cite{MT}.}

Equivariantly gauge-fixed Yang--Mills theories have two independent couplings, the gauge coupling $g$ and the
gauge-fixing parameter $\x$.   The latter can be traded for the ``longitudinal''
coupling $\tg$ defined by $\tg^2\equiv \x g^2$.
The couplings $g$ and $\tg$ are both asymptotically free \cite{GSb}.
In this article we will argue that the phase diagram
in the $g$--$\tg$ plane is non-trivial, with potentially interesting physical
consequences, despite the invariance theorem.

We will limit ourselves to a study of $SU(2)$ Yang--Mills theory,
equivariantly gauge-fixed to a $U(1)$ gauge theory.   In the framework
of equivariant gauge fixing,
$SU(2)$ gauge transformations divide into the gauge transformations
contained in
the $U(1)$ subgroup, which will remain unfixed, and the remainder, which
live in the $SU(2)/U(1)$ coset, and which will be gauge-fixed such that the
resulting gauge theory is invariant under eBRST symmetry.

While both $g$ and $\tg$ grow towards the infrared because
of asymptotic freedom,
the renormalization-group equation for $\tg$ depends on $g$ in such
a way that $\tg$ has to become strong in the infrared along with $g$.
This can happen in two different ways.   One scenario is that both couplings
become strong simultaneously, going into the infrared.
 The other scenario is
that $\tg$ becomes strong first, while $g$ remains relatively weak.
A third scenario, in which $\tg$ remains weak while $g$ becomes strong
is excluded by the renormalization-group analysis.

Which of the two possible scenarios is realized depends on the relative size of the couplings
at the cutoff.    Based as it is on solving the one-loop renormalization-group
equations, we do not know what actually happens when one or both
couplings have become strong.
However, in the case of the second scenario, we do know that the physics of the longitudinal sector, which is governed by $\tg$, becomes strongly coupled first, while $g$ still remains weak, keeping the transverse sector still perturbative.

A natural framework for studying the second scenario is the so-called
reduced model, which corresponds to the $g=0$ boundary of the phase diagram.
The reduced model is constructed as follows.  First, we perform a
gauge transformation, so that the  gauge transformations
become explicit as an $SU(2)$-valued scalar field in the path integral defining the theory.   The action depends on this field because the gauge-fixing
and ghost terms are not invariant under $SU(2)/U(1)$ gauge transformations.
Then, we turn off the transverse gauge fields by setting $g=0$.
The result is a theory of an $SU(2)$-valued scalar field coupled to the ghosts.
The reduced model has one coupling,
$\tg$, which is asymptotically free.
Since only the coset $SU(2)/U(1)$ was gauge-fixed, the remaining $U(1)$ remains a gauge symmetry, also in the reduced
model.   The growth of $\tg$ towards the infrared
suggests that dynamical symmetry breaking might take place at low energy
in the reduced model,
and it is this question, along with its consequences, that we investigate in this article.

The reduced model is invariant under $U(1)_L\times SU(2)_R$ transformations,
where $U(1)_L$ is a local symmetry and $SU(2)_R$ is a global symmetry.
The spontaneous breaking of the global $SU(2)_R$ symmetry
can be studied within the reduced model by standard methods.
Using a combination of strong-coupling and mean-field techniques,
we will find that a phase in which $SU(2)_R$ is broken to
an abelian subgroup $U(1)_R$ may appear at large $\tg$.

While the existence of the phase with broken symmetry will have to be studied
in the future using more reliable methods than mean field,
it is interesting to address the
consequences of such a phase for the full theory.   Since the
reduced model corresponds to the $g=0$ boundary of the phase diagram,
we recover the full theory by turning $g$ back on.   Our aim is then
to see whether any dynamical
symmetry breaking that occurs in the reduced model might have consequences
for the physics of the transverse degrees of freedom.

Naively, one expects this not to be the case.   According to the standard
paradigm, physics in the transverse sector should be independent of the
choice of gauge; therefore it should be blind to any symmetry breaking
in the longitudinal sector that may have occurred in the reduced model.
   Indeed, in a finite
volume, this is true:   The invariance theorem introduced
above guarantees that gauge
invariant correlation functions in the full theory are independent of $\tg$,
and equal to those in the unfixed theory \cite{gfx}.%
\footnote{The latter, of course, is well-defined
in the compact formulation on the lattice.}

However, this may not be the whole story.
Let us introduce a small breaking of eBRST symmetry
(which we will generically refer to as a ``seed'') into the theory.
We take the infinite-volume limit first, and then turn off the seed.
With this order of limits, the invariance theorem ceases to hold.
There are two possible outcomes.
The first possibility is that the consequences of the theorem are recovered
when the seed is eventually taken to zero.  Then the transverse sector of the
full theory has the same physics as that of the unfixed gauge theory.
Another possibility is that the dynamics of the longitudinal sector carries over to the transverse sector, thus uncovering a new phase in the full theory with
physical consequences.

In the reduced model there are local order parameters that signal
the spontaneous breaking of the global $SU(2)_R$ symmetry.
Once we turn back on the transverse gauge fields, which is equivalent
to gauging the $SU(2)_R$ symmetry,
no local order parameter is available any more to monitor its breaking
\cite{Elitzur}.   Nevertheless, the phases of the theory can still
be rigorously distinguished.   The familiar confining phase of an
unfixed $SU(2)$ Yang--Mills theory is characterized by a mass gap.
By contrast, if the spontaneous breaking of $SU(2)_R$
in the longitudinal sector carries over to the transverse sector,
the gauge fields in the $SU(2)/U(1)$ coset, which we will refer to as the $W$
fields,\footnote{In more physical terms,
the gauge fields which are charged under the unfixed $U(1)$.}
become massive.  The third gauge field stays massless,
and the long-distance physics is that of an abelian theory.
Therefore, phases corresponding to the two different possibilities can be
distinguished by the presence or absence of a mass gap \cite{GG,LS}.

As in any asymptotically-free theory, the infrared physics
is non-perturbative.  The only proven method for the investigation
of this non-perturbative dynamics is numerical, through
Monte-Carlo evaluation of correlation functions in the lattice-regularized
version of the theory.   However, a number of analytic techniques is available
which, although heuristic and/or generally not applicable near the continuum limit, can help in gaining insight into the nature of the phase diagram.   Here, we begin
with an analytic study of parts of the phase diagram through a combination of strong-coupling and mean-field techniques.
The motivation for this exploratory study is two-fold:   First, it is interesting to use these
techniques to form an idea of what the physics of equivariantly gauge-fixed
Yang--Mills theories might look like.
At a more concrete level, the results of such an analytic study provide
a framework for numerical studies, by identifying the important observables
as well as the properties of a phase in which the (consequences of the) invariance
theorem are avoided.

In Sec.~\ref{Review} we review the definition of the $SU(2)/U(1)$
 equivariantly gauge-fixed  Yang--Mills theory,
listing all symmetries, and including a definition of
the reduced model.  This is followed in Sec.~\ref{symmetry} by a discussion of
possible patterns of symmetry breaking.
As a corollary of the invariance theorem one can show that the reduced model
is a topological field theory, with a partition function that is a pure number,
independent of the parameters of the action.
We demonstrate, through a toy model,
that a topological field theory can nevertheless accommodate a non-trivial
effective potential for an order parameter, and  phases with a broken
symmetry.  In Sec.~\ref{SC} we derive the
effective action for the reduced model to order $1/\tg^2$ by integrating
out the ghosts in a strong-coupling expansion, to which we then apply
mean-field techniques in Sec.~\ref{MF}.   We find that spontaneous
symmetry breaking from $SU(2)_R$ to $U(1)_R$ occurs in mean field.
Turning on a weakly-coupled transversal field,
we derive an expression for the massive $W$ propagator. 
We then use these results, augmented by additional
considerations, to discuss the possible phase diagram of the theory in
Sec.~\ref{phase}, and conclude in Sec.~\ref{Conclusion}.
There are several appendices.
In App.~\ref{special} we generalize the invariance theorem
to the case where $G=SU(N)$ and $H$ is any maximal subgroup of $G$.
In App.~\ref{Toydetails} we prove a result pertaining to the toy model
of Sec.~\ref{symmetry}. Appendix~\ref{Group} summarizes some group integrals.
Appendix~\ref{Alternative} discusses an alternative mean-field analysis,
and App.~\ref{PT} proves a perturbative result referred to in the main
text.
Throughout this article we will set the lattice spacing equal to one.

\section{\label{Review} Equivariantly gauge-fixed Yang--Mills theory}
In this section we define the theory of interest, the $SU(2)$ equivariantly gauge-fixed Yang--Mills theory.
In this case, the only non-trivial subgroup is $H=U(1)$.
We will be brief in this section; for a general discussion of the construction
of equivariantly gauge-fixed theories, we refer to Ref.~\cite{gfx}.

\subsection{\label{vector} Vector picture}
We begin with the field content of our theory.   The lattice gauge field
consists of $SU(2)$-valued link variables $U_{x,\m}$, and can be written in
terms of a Lie-algebra valued field $V_{x,\m}$ through
\begin{equation}
\label{V}
U_{x,\m}=e^{iV_{x,\m}}\ ,\qquad V_{x,\m}=V_{x,\m,i}\,\frac{\t_i}{2}\ ,
\end{equation}
with $\t_i$, $i=1,2,3$, the Pauli matrices.
The gauge-fixing action depends on ghost and anti-ghost fields, $C_x$ and
$\bc_x$, and on a real auxiliary field $b_x$.   Since we are
only fixing the coset $SU(2)/U(1)$, these fields live in the corresponding
coset of the Lie algebra.   In components, the ghost field thus reads
\begin{equation}
\label{ghost}
C_x = C_{x,1}\frac{\t_1}{2}+C_{x,2}\frac{\t_2}{2}\ ,
\end{equation}
with similar expansions for the anti-ghost and auxiliary fields.

The lattice action defining the theory is given by
\begin{subequations}
\label{action}
\begin{eqnarray}
S&=&S_{\rm gauge}+S_{\rm gf}\ ,\label{actiona}\\
S_{\rm gf}&=&\tr\sum_x\Biggl(\tg^2 b_x^2-2i b_xD_\m^-\cw_{x,\m}\ ,\label{actionb}\\
&&+
\frac{1}{2}[\t_3,D_\m^+C_x][U_{x,\m}\t_3U_{x,\m}^\dagger,D_\m^+\bc_x]
\nonumber\\
&&+i\,\cw_{x,\m}\{\bc_x,D_\m^+C_x\}+\tg^2\{C_x,\bc_x\}^2\Biggr)\ .
\nonumber
\end{eqnarray}
\end{subequations}
Here $S_{\rm gauge}$ is any gauge-invariant lattice action constructed
from the link variables $U_{x,\m}$ that reduces in the classical continuum limit
to $\frac{1}{2g^2}\int d^4x\,\tr(F_{\m\n}^2)$, where
\begin{equation}
\label{fs}
F_{\m\n}=\partial_\m V_\n-\partial_\n V_\m+i[V_\m,V_\n]\ ,
\end{equation}
is the corresponding Lie-algebra valued continuum field strength,
and we have used the same notation for the continuum
gauge field as in Eq.~(\ref{V}).   The lattice vector potential can be decomposed as
\begin{eqnarray}
\label{decomp}
V_{x,\m}&=&W_{x,\m}+A_{x,\m}\,\frac{\t_3}{2}\ ,\\
W_{x,\m}&=&W_{x,\m,1}\,\frac{\t_1}{2}+W_{x,\m,2}\,\frac{\t_2}{2}\ ,\nonumber
\end{eqnarray}
with $W_{\m,1}$ and $W_{\m,2}$ the coset fields, and $A_\m$ the $U(1)$ field.
In Eq.~(\ref{actionb}), we have used the definitions
\begin{subequations}
\label{defs}
\begin{eqnarray}
\cw_{x,\m}&=&-\frac{i}{4}[U_{x,\m}\t_3U_{x,\m}^\dagger,\t_3]\ ,\label{defsa}\\
D_\m^+\Phi_x&=&U_{x,\m}\Phi_{x+\hm}U_{x,\m}^\dagger-\Phi_x\ ,\label{defsb}\\
D_\m^-\Phi_x&=&\Phi_x-U^\dagger_{x-\hm,\m}\Phi_{x-\m}U_{x-\hm,\m}\ ,\label{defsc}
\end{eqnarray}
\end{subequations}
where $\Phi$ is any Lie-algebra valued field.
It follows from Eq.~(\ref{defsa}) that
also $\cw_{x,\m}$ has components only in the coset.
In the classical continuum limit $\cw_{x,\m}\to W_{x,\m}$, and
$D_\m^-\cw_{x,\m}\to D_\m(A)W_\m$, with
$D_\m(A)$ the $U(1)$ covariant derivative.  Indeed, we see that this is a
gauge-fixing condition for the coset gauge field $W$, which leaves
the $U(1)$ gauge invariance intact.
A gauge-fixing parameter $\x$ may be defined by identifying
\begin{equation}
\label{gaugepar}
\tg^2\equiv\x g^2\ .
\end{equation}

Under $U(1)$ gauge transformations, $h_x=e^{i\theta_x\t_3/2}$, all coset fields
($\cw_{x,\m}$, $C_x$, $\bc_x$ and $b_x$) transform as
\begin{equation}
\label{cosetu1}
\Phi_x\to h_x\Phi_x h^\dagger_x\ ,
\end{equation}
and the action is thus $U(1)$ gauge invariant.   The remaining $SU(2)$
gauge transformations are broken by $S_{\rm gf}$.   Instead, the action
is invariant under the eBRST transformations
\begin{eqnarray}
\label{ebrst}
sU_{x,\m}&=&i(D_\m^+C)_xU_{x,\m}=i\left(U_{x,\m}C_{x+\hm}-
C_xU_{x,\m}\right)\ ,\\
sC_x&=&0\ ,\nonumber\\
s\bc_x&=&-ib_x\ ,\nonumber\\
sb_x&=&i[C^2_x,\bc_x]\ .\nonumber
\end{eqnarray}
Normally, one would expect the ghost field to transform as $sC=-iC^2$,
but since $C^2$ points in the $\t_3$ direction, this is not allowed.
This change in the transformation rule of $C_x$ modifies the nilpotency property of $s$:  Instead of $s^2=0$, we now have
\begin{equation}
\label{nilpotent}
s^2(\mbox{field}) = \d_X(\mbox{field})\ ,\qquad X=iC^2\ ,
\end{equation}
where $\d_X$ is an infinitesimal $U(1)$ gauge transformation with
parameter $X$.   It follows that $s$ is nilpotent when acting on operators
which are invariant under $U(1)$ gauge transformations.

The action breaks not only the local, but also the global $SU(2)$ symmetry.
Apart from the $U(1)$ subgroup, the action is invariant under a discrete
subgroup $\ts_2$ generated by
\begin{equation}
\label{spg}
\tP_\a=e^{i\p\t_\a/2}=i\t_\a\ ,\quad \a=1,2\ ,
\end{equation}
under which
\begin{equation}
\label{Uspg}
U_{x,\m}\to\tP_\a U_{x,\m}\tP_\a^\dagger\ ,
\end{equation}
with the other fields all transforming in the same way.

In addition, the
action is invariant under a symmetry of the ghost sector
that we will refer to as Schaden symmetry \cite{sdn2}, generated by
\begin{equation}
\label{schaden1}
\P_+=C_\a\frac{\d}{\d\bc_\a}\ ,\qquad \P_-=\bc_\a\frac{\d}{\d C_\a}\ ,
\end{equation}
and
\begin{equation}
\label{schaden2}
\Pi_3=\frac{1}{2}\left(C_\a\frac{\d}{\d C_\a}-\bc_\a\frac{\d}{\d\bc_\a}\right)\ ,
\end{equation}
where the latter generates the $U(1)$ ghost-number symmetry.  These
three generators together generate the group $SL(2,R)$.

The construction described above applies to a more general class
of theories, in which $G=SU(N)$, and $H$ is any maximal subgroup of $SU(N)$.
This class of theories is introduced in App.~\ref{special},
where we also observe that the invariance theorem applies
to all theories in this class.   For such theories the gauge-fixing action,
generalizing Eq.~(\ref{actionb}), and defined in App.~\ref{special}, is the most general possible form.   In particular, all such theories have only
two couplings, $g$ and $\tg$.

Finally, for theories in this class ``flip'' symmetry, defined
in Ref.~\cite{gfx}, takes a simple form.   It transforms only the ghost and
anti-ghost fields, according to $C\to\bc$,
$\bc\to -C$, while leaving all other fields invariant.   Applying flip
symmetry to Eq.~(\ref{ebrst}), it follows that the action is also invariant under an
anti-eBRST symmetry $\ebb$.\footnote{While other
equivariantly gauge-fixed theories may have flip symmetry and anti-eBRST symmetry, these symmetries, as well as the eBRST symmetry~(\ref{ebrst}), take a more complicated form in general \cite{gfx}.}   It is then straightforward to show that
\begin{eqnarray}
\label{aebrst}
\ebb^2({\rm field})&=&\d_\bX({\rm field})\ ,\qquad\bX=i\bc^2\ ,\\
\{s,\ebb\}({\rm field})&=&\d_\tX({\rm field})\ ,\qquad\tX=i\{C,\bc\}\ ,\nonumber
\end{eqnarray}
analogous to Eq.~(\ref{nilpotent}).

\subsection{\label{Higgs} Higgs picture and reduced model}
The action~(\ref{action}) is not gauge invariant (except under the $U(1)$
subgroup), and the longitudinal part of the gauge field, or equivalently,
 the gauge degrees
of freedom thus couple to the other fields in the theory.
The standard paradigm is that $SU(2)$ gauge invariant correlators remain the
same as in the unfixed theory; but it is precisely the aim of this article to
investigate whether this is indeed true everywhere in the phase diagram.

The gauge degrees of freedom can be exposed
by carrying out a gauge transformation on the theory defined by Eq.~(\ref{action}).
In technical terms, the ``Higgs--St\"uckelberg'' picture, or, for short,
the ``Higgs'' picture of our theory is obtained by making the replacement
\begin{equation}
\label{gt}
U_{x,\m} \to U_{x,\m}^\phi \equiv \phi_x U_{x,\m}\phi^\dagger_{x+\hm}
\end{equation}
in Eq.~(\ref{action}), and integrating over the $SU(2)$ valued field $\phi_x$.
Explicitly,
\begin{subequations}
\label{Zvh}
\begin{eqnarray}
  Z &=& \int \cd U \cd b \cd C \cd \bc \,
  \exp\Big(-S_{\rm gauge}(U) -S_{\rm gf}(U,b,C,\bc) \Big) \hspace{10ex}
\label{Zvha}\\
  &=& \int \cd U \, \exp\Big(-S_{\rm gauge}(U)\Big)\, \tZ(U) \,,
\label{Zvhb}
\end{eqnarray}
\end{subequations}
where
\begin{equation}
  \tZ(U) = \int \cd\phi \cd b \cd C \cd \bc \,
  \exp\Big(-S_{\rm gf}(U^\phi,b,C,\bc) \Big) \,.
\label{Zred}
\end{equation}
Equation~(\ref{Zvha}) is the partition function in the vector picture,
introduced in the previous subsection.  Equation~(\ref{Zvhb}), together
with Eq.~(\ref{Zred}), gives the Higgs-picture partition
function.  Except for the integration over the gauge field itself,
we have absorbed the integration over all other fields,
including the gauge transformations described by $\phi_x$,
into Eq.~(\ref{Zred}).
The motivation for doing so will become clear shortly.
Since $S_{\rm gauge}$ is gauge invariant,
only $S_{\rm gf}$ is effected by the gauge transformation.
One can always go back to the vector picture
by making the field redefinition
$U_{x,\m}\to\phi_x^\dagger U_{x,\m}\phi_{x+\hm}$ which,
when substituted into the right-hand side of Eq.~(\ref{Zred})
eliminates all dependence on $\phi_x$ from the action.

The symmetry structure of the action is modified in the Higgs picture.
Equivariant BRST transformations are as in Eq.~(\ref{ebrst}),
except that now $U_{x,\m}$ is invariant, while the Higgs--St\"uckelberg
field $\phi_x$ transforms as
\begin{equation}
\label{ebrstphi}
s\phi_x=-iC_x\phi_x\ .
\end{equation}
This makes the combination~(\ref{gt}) transform as in the first
line of Eq.~(\ref{ebrst}), thus maintaining the eBRST invariance of the action.

An inspection of Eq.~(\ref{gt}) shows that the theory is now invariant under an
extra $SU(2)$ gauge symmetry,
\begin{eqnarray}
\label{su2r}
U_{x,\m}&\to& g_x U_{x,\m}g^\dagger_{x+\hm}\ ,\\
\phi_x&\to&h_x\phi_x g^\dagger_x\ ,\nonumber
\end{eqnarray}
in which $g_x\in SU(2)$.   In Eq.~(\ref{su2r}) we also show how these fields
transform under the $U(1)$ gauge symmetry~(\ref{cosetu1}) in the Higgs
picture.  The full gauge symmetry of the action in the
Higgs picture is thus the group $U(1)_L\times SU(2)_R$, where we
used the labels $L$ and $R$ to indicate on which side of the field $\phi_x$
these transformations act.

The action $S_{\rm gauge}$, with gauge coupling $g$,
participates only in the dynamics of the
transverse part of the gauge field, while $S_{\rm gf}$, with coupling $\tg$,
also governs the dynamics of the rest of the degrees of freedom.
Here, we are interested in possible strong $\tg$-dynamics while keeping the
gauge coupling $g$ perturbative.   It then makes sense to first consider
the limiting case $g=0$, which freezes out the transverse part of the
gauge field described in the Higgs picture by $U_{x,\m}$.
In this limit, the theory is thus defined by $S_{\rm gf}$, in which first
the replacement~(\ref{gt}) has been made, followed by setting $U_{x,\m}=I$.

We will refer to this simplified version of the theory as the reduced
model.  In short, the reduced model's action is given by Eq.~(\ref{actionb}),
with the replacement $U_{x,\m}\to\phi_x\phi_{x+\hm}^\dagger$.
Its partition function is $\Zred = \tZ(I)$.
This theory is invariant under eBRST with the rule~(\ref{ebrstphi}), as well
as under the group $U(1)_L\times SU(2)_R$, where now
$SU(2)_R$ has become a global symmetry.%
\footnote{ The full theory, in the Higgs picture, can be recovered
  from the reduced model by introducing an $SU(2)$ gauge field and promoting
  $SU(2)_R$ to a local symmetry.}
The role of the local $U(1)_L$ symmetry is to effectively make $\phi_x$
take values in the coset $SU(2)/U(1)$, with two fields $\phi_x$ and $\phi'_x$
being in the same equivalence class if they differ only by a $U(1)_L$ gauge
transformation.

A key question we will address in this article is how the global
$SU(2)_R$ symmetry of the reduced model is realized as a function of $\tg$.

\section{\label{symmetry} Patterns of symmetry breaking}
The invariance theorem states that finite-volume expectation values
of gauge-invariant operators in the eBRST gauge-fixed theory
are exactly equal to their values in the unfixed theory \cite{gfx}.
The partition function $Z$ is equal to that of the unfixed theory
up to a non-vanishing multiplicative constant.
Briefly, the proof works by noting that $S_{\rm gf}$ can be written
as an eBRST variation, \seef Eq.~(\ref{lattgf}).  Multiplying
the first term on the right-hand side of Eq.~(\ref{lattgf}) by a parameter $t$,
it follows that $\partial Z/\partial t$ vanishes, because it is
the expectation value of an eBRST variation.
This conclusion generalizes to expectation values of gauge invariant
operators.  In the limit $t\to 0$,
the first term on the right-hand side of Eq.~(\ref{lattgf}) drops out,
while the remaining terms depend only on the ghost-sector fields.
The partition function $Z$ of Eq.~(\ref{Zvh}) is therefore a product
of the original gauge-invariant partition function,
and a decoupled partition function for the ghost-sector fields.
Moreover, the latter is a product of decoupled integrals
at each lattice site, which can be shown to yield
a non-zero constant \cite{gfx}.

In the Higgs picture, the invariance theorem applies
even if we keep the gauge field $U_{x,\m}$ external, because
$U_{x,\m}$ does not transform under eBRST in this case.
It follows that the partition function $\tZ(U)$
of Eq.~(\ref{Zred}) defines a topological field theory: While the action $S_{\rm gf}$
depends on $\tg$, the partition function $\tZ(U)$ does not, and
it is also independent of the external gauge field $U_{x,\m}$.
It is just a pure number.

The existence of the invariance theorem would lead one to expect that
if any spontaneous symmetry breaking took place in the reduced model,
this could not possibly have any effect on the physics of
the transverse gauge fields.  This state of affairs would be
in accordance with the standard paradigm that,
whatever gauge-fixing procedure one applies to a Yang--Mills theory,
the transverse dynamics is unchanged.  Indeed, to the extent
that exact eBRST invariance is maintained in any finite volume,
the invariance theorem applies, with the anticipated consequences.

Furthermore, one can question the proposition that the reduced model
could have any non-trivial phase structure at all.
A phase transition occurs when the effective potential for some
order parameter develops a new global minimum, as a function of
the couplings of the theory.  For this to happen,
the effective potential should depend on the couplings (here only $\tg$)
in a non-trivial way in the first place.
But the reduced model is a topological field
theory, with a partition function that is independent of the
parameter $\tg$.  The question arises whether such a partition function
can nevertheless accommodate a dynamics that generates
a non-trivial effective potential and gives rise to
spontaneous symmetry breaking.

In the following subsection we will illustrate and
address these questions in a toy model
that shares the most relevant features of the reduced model.
The toy model is a zero-dimensional ``field theory,'' \ie, an ordinary
integral.  This exercise will show that a topological (field) theory
can in principle break a symmetry spontaneously.
Encouraged by this result we return in Sec.~\ref{patterns} to the
equivariantly gauge-fixed theory
itself, and study patterns of spontaneous symmetry breaking
that might be triggered by strong $\tg$ dynamics.

\subsection{\label{Toy} Toy model}
In this subsection we demonstrate through a toy model that
global symmetries can, in principle, break spontaneously
in a theory satisfying a similar invariance theorem.
Our toy model is a zero-dimensional field theory with an exact BRST-type
invariance, as well as a discrete $Z_2$ symmetry.
In a sense that will be clarified below, the $Z_2$ symmetry
gets broken spontaneously, whereas BRST symmetry remains unbroken.

The model contains two real degrees of freedom, $\phi$ and $b$, and a
ghost, anti-ghost pair $c$, $\bct$.
The nilpotent BRST transformation $s$ acts as follows
\begin{eqnarray}
\label{tbrst}
s\phi&=&c\ ,\\
sc&=&0\ ,\nonumber\\
s\bct&=&ib\ ,\nonumber\\
sb&=&0\ .\nonumber
\end{eqnarray}
We choose the action to be%
\footnote{
  This action is reminiscent of the supersymmetric Wess--Zumino model,
except here we can use real bosonic fields.
}
\begin{equation}
\label{taction}
S=s\left(-i\bct b+\bct f(\phi)\right)=b^2+ibf(\phi)-\bct f'(\phi)c\ ,
\end{equation}
where
$f'(\phi) = \partial f/\partial\phi$.  We assume that,
as $\phi$ ranges from $-\infty$ to $+\infty$, so does $f(\phi)$.

BRST invariance of the toy model, together with the fact that the action
is a BRST variation, leads to an invariance theorem:
The partition function $Z$ is independent of any parameters inside $f$.
Indeed,
\begin{subequations}
\label{tZtoy}
\begin{eqnarray}
Z&\equiv&\frac{1}{2\p}\int_{-\infty}^\infty d\phi\int_{-\infty}^\infty db
\int dcd\bct\,e^{-\left(b^2+ibf(\phi)-\bct f'(\phi)c\right)}\label{tZtoya}\\
&=&\frac{1}{2\sqrt{\p}}\int_{-\infty}^\infty d\phi\,f'(\phi)\,e^{-f^2(\phi)/4}=1\ .
\label{tZtoyb}
\end{eqnarray}
\end{subequations}
The only reservation is that the assumption we have made about the asymptotic
behavior of $f(\phi)$ must be respected.  Therefore, if $f(\phi)$
is a polynomial, it must be of odd degree, and
the coefficient of the highest power must be positive.
The coefficients of the rest of the polynomial can then be changed at will,
without altering the value of $Z$.

Now, let us consider a concrete example.   We choose
\begin{equation}
\label{tf}
f(\phi)=\frac{1}{\l}\left(\phi^3-v^2\phi\right)\ ,
\end{equation}
with $\l$ and $v^2$ two real parameters.
In addition to the BRST symmetry~(\ref{tbrst}), this choice leads to
invariance under  a discrete $Z_2$ symmetry that flips the signs of $\phi$
and $b$.

The ``classical potential''  $f^2(\phi)/4$ has a single minimum
at $\phi=0$ when $v^2<0$.  But, for $v^2>0$,
additional degenerate minima appear at $\phi=\pm v$.
These new minima are not invariant under the $Z_2$ discrete symmetry.
Were we dealing with a true field theory, we would
observe a spontaneous breaking of the $Z_2$ symmetry by introducing
a ``seed'' that breaks this symmetry explicitly into the finite-volume action,
and turning it off after the thermodynamic limit has been taken.\footnote{
  As an example, a similar discrete symmetry undergoes spontaneous
  breaking in the Ising model in $d\ge 2$ euclidean dimensions.
}
Choosing the seed to be
\begin{equation}
\label{tseed}
S_{\rm seed}= -\e\phi \ ,
\end{equation}
the minimum $\phi=v>0$ would be selected for $\e>0$,
while $\phi=-v$ would be selected for $\e<0$.

Being a zero-dimensional field theory, in the toy model there is no
thermodynamic limit to be taken.
Strictly speaking, we must always keep $\e$ non-zero
to maintain a preference for
a particular saddle point with broken symmetry.
The situation is more favorable in perturbation theory.
As usual, when the coupling constant $\l$ is small,
we may select any one of the saddle points and develop
a perturbative expansion around it.%
\footnote{
  As usual, the expansion in $\l$ is an asymptotic expansion.
}
Within this framework, a seed is not introduced.
We will denote perturbatively calculated quantities obtained this way
with a subscript that identifies the saddle point.

We first consider the perturbative expansion around the minimum at $\phi=0$.
Setting $\phi=\l\psi$ and integrating over $\psi$ we find
\begin{subequations}
\label{tsaddle2}
\begin{eqnarray}
Z_0&=&-1\ ,\qquad\mbox{to all orders,}
\label{tsaddle2a}\\
\langle\phi\rangle_0&=&0\ ,\qquad\mbox{to all orders.}
\label{tsaddle2b}
\end{eqnarray}
\end{subequations}
An explicit calculation of $Z_0$ can be found in App.~\ref{Toydetails}.
It is easily seen that the vanishing result for $\svev{\phi}_0$ generalizes
to every order parameter for the $Z_2$ symmetry, and therefore
this symmetry is not broken spontaneously at the $\phi=0$ saddle point.

For the other two saddle points we expand $\phi=\pm v+\l\psi$.
We find
\begin{subequations}
\label{tsaddle}
\begin{eqnarray}
Z_{\pm v} &=& 1\ ,\qquad\mbox{to all orders,}
\label{tsaddlea}\\
\langle\phi\rangle_{\pm v} &=& \pm v\left(1-\frac{3}{4}\frac{\l^2}{v^6}
+O\left((\l^2/v^6)^2\right)\right)\ ,
\label{tsaddleb}
\end{eqnarray}
\end{subequations}
showing that indeed the $Z_2$ symmetry is broken spontaneously
at these saddle points.  Of course, summing over all
the saddle points (for vanishing seed) will restore the symmetry.
We will return to the proof of Eq.~(\ref{tsaddlea}) below.

Let us now consider how BRST symmetry is realized at the $\phi=\pm v$
saddle points.  While $\phi$ is not invariant under BRST,
an expectation value $\svev{\phi}_{\pm v}\ne 0$ does not imply the breaking
of BRST symmetry, because $\phi$ is not the BRST variation of anything;
spontaneous BRST symmetry breaking would be signaled by
$\svev{sX}_{\pm v} \ne 0$ for some $X$.
We can in fact prove that BRST symmetry is not broken within the perturbative
expansion around any saddle point.  Taking for example the
saddle point $\phi=v$ we first expand $\phi=v+\l\psi$, and then define
the BRST rule to be $s\psi=c/\l$.  This rule is clearly consistent
with Eq.~(\ref{tbrst}).  It follows that the perturbative expansion
around the saddle point respects BRST symmetry.
We may derive BRST Ward identities in the usual way,
obtaining that $\svev{sX}_v=0$ for any $X$.  The same reasoning
applies at the other saddle points.

An interesting corollary is that the invariance theorem applies within
the perturbative expansion around each saddle point of the toy model.
The saddle-point partition functions $Z_0$ and $Z_{\pm v}$ must
therefore be independent, in particular, of the coupling constant $\l$.
Since these partition functions
are intrinsically defined within a perturbative expansion in $\l$,
it follows that their tree-level values receive no corrections
to all orders in $\l$.  The tree-level values can easily be found,
leading to Eqs.~(\ref{tsaddle2a}) and~(\ref{tsaddlea}).%
\footnote{
Note that the sum over the three saddle points $Z_{+v}+Z_{-v}+Z_0$
reproduces the exact partition function~(\ref{tZtoy}).
As already mentioned, an explicit verification of Eq.~(\ref{tsaddle2a})
to all orders is given in App.~\ref{Toydetails}.
}

While the saddle-point partition functions
are independent of $\l$ and $v$, our explicit calculation
shows that observables, such as $\langle\phi\rangle_{\pm v}$,
depend on these parameters in a non-trivial way.
The technical reason is easily understood.
We may calculate $\svev{\phi}_{\pm v}$ by augmenting the action
with a source term, $-J\psi$, and taking the derivative of
$W_{\pm v}(J) = -\log Z_{\pm v}(J)$ with respect to $J$.
The source term is recognized as nothing but the seed, Eq.~(\ref{tseed}),
with $\e\to J$.  This term is not invariant under BRST,
leading to the failure of the invariance theorem.
Hence, $W_{\pm v}(J)$ can depend non-trivially on $\l$ and $v$.
The same conclusion applies to the effective action $\G_{\pm v}(\phi_{\rm eff})$
obtained through a Legendre transformation, which, in the case of
the toy model, coincides with the effective potential for the order
parameter $\svev{\phi}_{\pm v}$.

The main lesson we have learned is that topological (field)
theories can accommodate non-trivial effective potentials
for order parameters, as well as symmetry-breaking saddle points.
Later on we will find evidence that similar conclusions hold
in the eBRST gauge-fixed theory that is the main subject of this paper.

\subsection{\label{patterns} Symmetry breaking in the equivariantly gauge-fixed theory}
In this subsection we consider a symmetry breaking pattern that
might be triggered by strong $\tg$ dynamics in the reduced model,
and address its implications for the full theory.
We are encouraged by the results of the previous subsection,
which show that spontaneous symmetry breaking can in principle take
place in a topological field theory.  Even though the reduced model's
partition function is independent of the coupling constant $\tg$
(and of the external gauge field, \seef\ Eq.~(\ref{Zred})),
this does not preclude the existence of an effective
potential with non-trivial minima for order parameters.

We will be interested in a symmetry breaking pattern
where the global $SU(2)_R$ is broken down to an abelian subgroup
$U(1)_R$, while eBRST symmetry remains unbroken.
A natural order parameter for this symmetry breaking is
the expectation value of
\begin{equation}
\label{A3}
\aop_x=\phi^\dagger_x\t_3\phi_x\ .
\end{equation}
This operator is invariant under the local $U(1)_L$ symmetry, as it should be.
It transforms in the adjoint representation of $SU(2)_R$, and therefore
a non-zero expectation value, $\vev{\aop}\ne 0$,
signals the symmetry breaking $SU(2)_R\to U(1)_R$.

While the operator $\aop$ is not invariant under eBRST, its
expectation value cannot serve as an order parameter for
eBRST symmetry breaking.
Again the reason is that $\aop$ is not the eBRST variation
of any other operator, and therefore it will not occur in any Ward identity
for eBRST symmetry.

Following standard practice, in order to study the anticipated
symmetry breaking pattern we add to the finite-volume action the term
\begin{equation}
\label{seed}
S_{\rm seed}=-h\sum_x\,\tr(\t_3 \aop_x) \ ,
\end{equation}
choosing $h>0$.  The limit $h \to 0$ is taken after the thermodynamic limit.
The seed tilts the effective potential for the order parameter,
and, in the broken phase, selects the vacuum state
where $\vev{\aop}$ points in the $\t_3$ direction.
Once the infinite-volume limit has been
taken, transitions between different vacua are kinematically suppressed.
Hence the vacuum state will remain that selected by the seed when
the latter is turned off.

As usual, we may obtain the order parameter $\vev{\aop}$
from the logarithmic derivative of the partition function with
respect to the external magnetic field $h$.  Once again,
the seed~(\ref{seed}) is not invariant under eBRST transformations,
leading to the failure of the invariance theorem for non-zero $h$.
Much like the toy model of Sec.~\ref{Toy},
this allows $\vev{\aop}$ to depend non-trivially
on the coupling $\tg$.

Next, we turn to the full theory, which corresponds to moving
away from the $g=0$ boundary into the $g>0$ phase diagram.
We will assume that the symmetry breaking discussed above takes
place in the reduced model.  For definiteness, anticipating
our analysis in subsequent sections,
we assume the existence of a critical value
$\tg_c$ such that, if we start from the strong-coupling limit,
the reduced model is in a symmetric phase for $\tg_c<\tg<\infty$,
while for $\tg<\tg_c$ we enter the phase in which $SU(2)_R$ is broken
down to $U(1)_R$.

In lattice Higgs models such as Ref.~\cite{LS},
no symmetry-breaking seed is needed once the global symmetry
is promoted to a local symmetry.  Here, too, a seed
that breaks the $SU(2)_R$ symmetry is not needed for $g>0$.
But, as explained above, in the equivariantly gauge-fixed theory
it is necessary to turn on a seed that breaks the eBRST symmetry
in order to probe any non-trivial phase structure.
Because of the invariance theorem, we are bound to recover the usual
confining physics everywhere in the $g>0$ phase diagram
if we did not introduce an eBRST-breaking seed.

Let us first consider what happens when we move into the phase diagram
starting from some $\tg>\tg_c$ in the symmetric phase.
The presence of the eBRST-breaking seed, and the ensuing failure of the invariance
theorem, imply that gauge invariant observables will be somewhat
distorted relative to the un-gauge fixed Yang--Mills theory.
However, in this region of the phase diagram we expect
to recover the consequences of the theorem in the thermodynamic limit,
and thus, that the theory is in the usual confining phase.

If, instead, we move into the $g>0$ phase diagram
starting at a point inside the broken phase of the reduced model,
it is in fact very natural
to find the full theory in a different phase resembling
the broken phase of the $SU(2)$ theory with an adjoint Higgs field \cite{GG,LS}.
In the reduced model, the symmetry breaking of $SU(2)_R$ down to $U(1)_R$
produces two Goldstone bosons.   In the full theory,
$SU(2)_R$ is promoted to a gauge symmetry (see Sec.~\ref{Higgs}).
Therefore, we would expect
a Higgs-like phenomenon: the two Goldstone bosons will turn into
the longitudinal modes of the $W$ fields which, in turn, will become massive.
Since the remaining unbroken gauge symmetry is abelian,
the gauge coupling will stop growing on larger distance scales,
and the third gauge field will stay massless.
We will refer to such a phase, if it exists, as a Coulomb phase.

As it turns out, the emergence of a $W$-boson mass term
is consistent with the preservation of eBRST symmetry
in the limit of vanishing seed,
provided that the ghost fields acquire the same mass.
Indeed, if we consider the (continuum) action
\begin{equation}
  S_{\rm m} = m^2 \int d^4x\, \tr (W_\m^2/2 + \bc C)\ ,
\label{Smass}
\end{equation}
we find that it is eBRST invariant on-shell,
\begin{eqnarray}
  s S_{\rm m} &=& m^2 \int d^4x\, \tr (W_\m D_\m(A)C  -ib C)
\label{sSmass}\\
  &=& -m^2 \int d^4x\, \tr \Big((D_\m(A)W_\m +ib)C\Big) = 0 \ ,
\nonumber
\end{eqnarray}
where we have integrated by parts and, in the last step,
used the auxiliary field's equation of motion.%
\footnote{
In view of the eBRST invariance of $S_{\rm m}$,
the question arises whether this mass term could be induced
in perturbation theory.  In App.~\ref{PT} we prove
that this is not the case.}

Finally, we recall that the couplings $g$ and $\tg$ are both
asymptotically free.
If a Coulomb phase indeed exists in the phase diagram of the full theory,
the most interesting question is whether this phase extends to the
gaussian fixed point $(g,\tg)=(0,0)$,
where the correlation lengths associated with both the transverse
and longitudinal sectors diverge in lattice units.
While we will not be able to answer
this question in this article, we will return to a further discussion of this
possibility, as well as a discussion of the renormalization-group evolution,
in Sec.~\ref{phase}.

\section{\label{SC} Effective theory at large $\tg$: integrating out the ghosts}
In the next two sections, we will use a combination of strong-coupling
and mean-field techniques in order to investigate the phase diagram of
the reduced model.   The reduced model has only one coupling, $\tg$,
and its phase diagram is thus one-dimensional.   This phase diagram
corresponds to the $g=0$ boundary of the two-dimensional phase
diagram of the full theory.

In this section, we will integrate over the ghost and anti-ghost fields in
a $1/\tg^2$ expansion, thus deriving an effective action for the
gauge field $U_{x,\m}$, to leading order in this expansion.   In the next
section, we will first restrict ourselves to the reduced model by setting
$U_{x,\m}=\phi_x\phi_{x+\hm}^\dagger$, and
look for phase transitions in the reduced model by
applying mean-field methods to this effective action.
We will then investigate the implications of the mean-field results
for the full theory.

Integrating over the
auxiliary field $b_x$ in Eq.~(\ref{actionb}), we obtain the on-shell version of
the gauge-fixing action,
\begin{equation}
\label{onshell}
S_{\rm gf}^{\rm on}=\frac{1}{\tg^2}\,\tr\sum_{x,\m}(D_\m^-\cw_{x,\m})^2
+\tg^2\,\tr\sum_x\{C_x,\bc_x\}^2+\sum_{xy,\a\b}\bc_{x,\a}\O_{x,\a;y,\b}C_{y,\b}\ .
\end{equation}
The matrix $\O$, defining the bilinear ghost action,
has the following non-zero components:
\begin{eqnarray}
\label{omega}
\O_{x,\a;x,\b}&=&\frac{1}{2}\,\d_{\a\b}\,\tr\left(\t_3 U_{x-\hm,\m}\t_3 U^\dagger_{x-\hm,\m}+\t_3 U_{x,\m}\t_3 U^\dagger_{x,\m}\right)
\equiv\frac{1}{2}\,\d_{\a\b}\hO_{xx}\ ,\\
\O_{x,\a;x+\hm,\b}&=&\frac{1}{2}\left(\tr\left(U_{x,\m}\t_\a U^\dagger_{x,\m}\t_\b\right)-\d_{\a\b}\,\sum_\g\tr\left(U_{x,\m}\t_\g U^\dagger_{x,\m}\t_\g\right)\right)\ ,\nonumber\\
\O_{x+\hm,\a;x,\b}&=&\frac{1}{2}\left(\tr\left(\t_\a U_{x,\m}\t_\b U^\dagger_{x,\m}\right)-\d_{\a\b}\,\sum_\g\tr\left(U_{x,\m}\t_\g U^\dagger_{x,\m}\t_\g\right)\right)\ ,\nonumber
\end{eqnarray}
with coset indices $\a,\b,\g=1,2$, and a sum over $\m$ implied
in the first line.  It can be checked that $\O$ is symmetric and real.

We may now integrate over the ghost fields in a strong-coupling expansion in
$\tg$, using that for large $\tg$ the dominant term in Eq.~(\ref{onshell})
is the single-site four-ghost term.
We define the normalization of single-site ghost integrals by requiring that
\begin{equation}
\label{ghint}
\int[dCd\bc]C_\a\bc_\b C_\g\bc_\d=\e_{\a\g}\e_{\b\d}\ ,
\end{equation}
where $\e_{\a\b}$ is the two-dimensional epsilon tensor.   It is then
straightforward to calculate the effective action for $U_{x,\m}$ to order
$1/\tg^2$, and we find
\begin{equation}
\label{Seff}
S_{\rm eff}=\frac{1}{\tg^2}\sum_x\left(
\tr\left(D_\m^-\cw_{x,\m}\right)^2
-\frac{1}{4}\,\hO_{xx}^2\right)+O\left(\frac{1}{\tg^4}\right)\ .
\end{equation}
At this order, $S_{\rm eff}$ only depends on the diagonal elements of $\O$.

\section{\label{MF} Dynamical symmetry breaking: mean-field study}
Our next step is to transition to the reduced model by substituting
\begin{equation}
\label{Usub}
U_{x,\m}=\phi_x e^{igV_{x,\m}}\phi^\dagger_{x+\hm}
\end{equation}
into Eq.~(\ref{Seff}), keeping the transverse gauge field $V_{x,\m}$ external.
This turns $S_{\rm eff}$ into a scalar theory, to which we will apply a
mean-field analysis.   First, in Sec.~\ref{reduced}, we set $V_{x,\m}=0$
and study the phase diagram of the reduced model as a function of
$\tb=1/\tg^2$.  Starting from $\tb=0$, as we increase $\tb$
the mean-field analysis predicts a first-order transition
into a phase where $SU(2)_R$ is spontaneously broken to $U(1)_R$.
Then, in Sec.~\ref{W mass}, in order to probe the consequences of
the spontaneous symmetry breaking in the reduced model
for the transverse gauge fields, we expand the full theory
around the mean-field solution to order $g^2$.
We find that in the broken phase the $W$ fields pick up a mass, in
accordance with our general remarks in Sec.~\ref{symmetry}.
Some further remarks on the mean-field analysis that put it into
a broader context are collected in Sec.~\ref{consistency}.

\subsection{\label{reduced} The reduced model}
The reduced model is invariant under a local $U(1)$ symmetry and a
global $SU(2)$ symmetry,
\begin{equation}
\label{redsymm}
\phi_x\to h_x\phi_x g^\dagger\ ,
\end{equation}
with $h_x\in U(1)_L$ and $g\in SU(2)_R$, \seef Eq.~(\ref{su2r}).  We repeat
the definition of the order parameter introduced in Sec.~\ref{symmetry},
\begin{equation}
\label{Aphi}
\aop(\phi)=\phi^\dagger\t_3\phi\ .
\end{equation}
This operator is invariant under the local $U(1)_L$.
A non-zero expectation value for this
composite field signals the symmetry breaking of $SU(2)_R$ down to
$U(1)_R$.   Indeed, we may define a three-dimensional vector $\aop_i(\phi)$
through
\begin{equation}
\label{adjoint}
\aop(\phi)=\aop_i(\phi)\t_i\quad\Rightarrow\quad \aop_i(\phi)=\frac{1}{2}\,\tr(\aop(\phi)\t_i)\ .
\end{equation}
This unit vector transforms in the adjoint representation of $SU(2)_R$.
If it acquires a non-zero expectation value this breaks $SU(2)_R$,
while leaving unbroken the $U(1)_R$ subgroup corresponding to
rotations around this vector.

At order $\tb$, the effective action of the reduced model is obtained
by substituting Eq.~(\ref{Usub}) with $V_{x,\m}=0$ into Eq.~(\ref{Seff}).
As it turns out, this effective action can be expressed solely
in terms of $\aop(\phi)$.%
\footnote{For some details on the determination of $S_{\rm eff}(\aop)$ from
$S_{\rm eff}(\phi)$, see App.~\ref{Alternative}.}
We may thus rewrite the partition function of the effective theory
for the reduced model as
\begin{eqnarray}
\label{ZMF}
Z&=&\int \prod_x d_H\phi_x\;e^{-S_{\rm eff}(\phi)}\\
&=&\int \prod_x d_H\phi_x\;e^{-S_{\rm eff}(\phi)}\prod_x\int da_{i,x}\int_{-i\infty}^{i\infty} dh_{i,x}\;
e^{-\frac{1}{2}\sum_x\tr(H_x(A_x-\aop(\phi_x)))}\nonumber\\
&=&\prod_x\int da_{i,x}\int dh_{i,x}\;e^{-\sum_x\left(\frac{1}{2}\tr(H_x A_x)+S_{\rm eff}(A_x)-u(H_x)\right)}\ ,\nonumber
\end{eqnarray}
where $d_H\phi$ is the Haar measure on $SU(2)$,
\begin{equation}
\label{AH}
H=h_i\t_i\ ,\qquad A=a_i\t_i\ ,
\end{equation}
are new fields with unconstrained real components $h_i$ and $a_i$, and
\begin{equation}
\label{uH}
e^{u(H)}=\int d_H\phi\;e^{\frac{1}{2}\,\tr(H\phi^\dagger\t_3\phi)}\ .
\end{equation}
The latter integral can be calculated by making use of the invariance
under $SU(2)_R$.
We find
\begin{equation}
\label{MFint}
e^{u(H)}=\frac{\sinh{h}}{h}\ .
\end{equation}
where $h=|{\vec h}|$ is the length of the vector.  For some steps of the
calculation of this integral, we refer to App.~\ref{Group}.

So far, our manipulations were exact.   The mean-field solution is obtained
by calculating the partition function in the saddle-point approximation
\cite{DZ}.  To this end we minimize the free energy
\begin{equation}
\label{free}
f(H,A)=h_i a_i+s_{\rm eff}(A)-u(H)\ .
\end{equation}
We assume that the mean-field solution is translationally invariant,
so that $A$ and $H$ are $x$-independent.
This leads to the mean-field equations
\begin{subequations}
\label{mfeq}
\begin{eqnarray}
a_i &=& \frac{\partial u}{\partial h_i}=\frac{\partial u}{\partial h}
\frac{h_i}{h}\ ,
\label{mfeqa}\\
h_i &=& \rule{0ex}{4ex} -\frac{\partial s_{\rm eff}}{\partial a_i}\ ,
\label{mfeqb}
\end{eqnarray}
\end{subequations}
with
\begin{equation}
\label{SA}
s_{\rm eff}=\frac{1}{V}\,S_{\rm eff}=-\tb\left(\frac{d}{4}(1-2d)\,\tr A^2+d^2\left(\tr A^2\right)^2
+\frac{d^2}{2}\,\tr A^4\right)\ ,
\end{equation}
where we wrote the mean-field action in $d$ rather than four dimensions.
Equation~(\ref{mfeqa}) implies that the orientation of the vector
$\vec{a}$ will follow that of $\vec{h}$.  We may thus choose
$h_1=h_2=a_1=a_2=0$ and write $h_3=h$ and $a_3=a$.

\begin{figure}
\begin{center}
\includegraphics*[width=12cm]{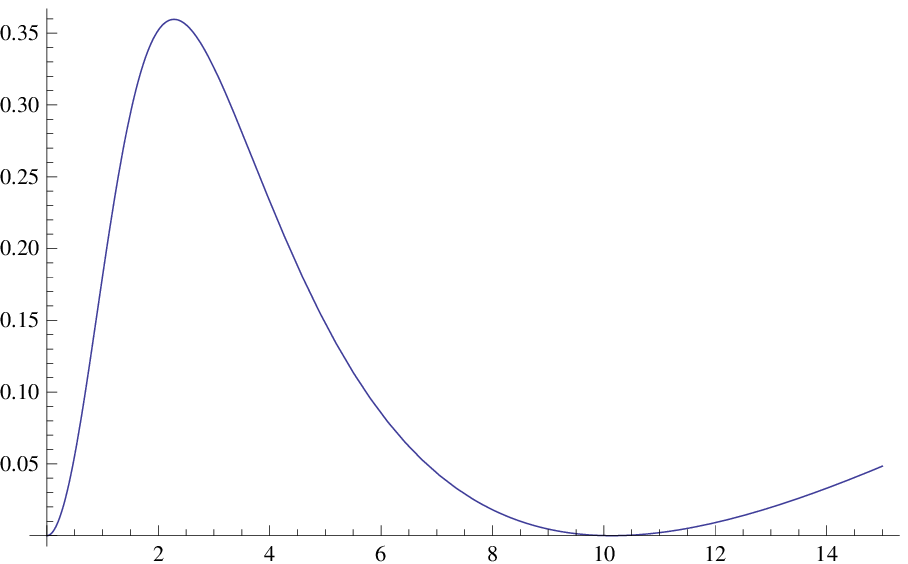}
\end{center}
\begin{quotation}
\floatcaption{free-mfb-new}%
{The free energy~(\ref{free}) as a function of $h$, with $d=4$ and
$\tb=0.04850$.}
\end{quotation}
\vspace*{-4ex}
\end{figure}

Solving for $a$ in terms of $h$, and substituting into Eq.~(\ref{free}),
we may plot the free energy as a function of $h$.
The free energy always has a zero at $h=0$.
For small $h$, the $d=4$ free energy may be expanded as
\begin{equation}
\label{freeexp}
  f(h) = \left( \frac{1}{6} + \frac{14}{9}\,\tb\right)h^2+O(h^4)\ .
\end{equation}
showing that the curvature at the origin is always positive.
This rules out a continuous phase transition.

For very small $\tb$, the free energy is positive
for $h>0$, and its absolute minimum is at $h=0$.
When $\tb$ increases, a new local minimum, $h_1(\tb)$, develops.
This minimum becomes degenerate with the global minimum at $h=0$
for $\tb=\tb_c=0.04850$.   The free energy for $\tb=\tb_c$
is shown in Fig.~\ref{free-mfb-new}.  The non-trivial minimum is at
$h=h_1(\tb_c)=10.084$, and the order parameter at this minimum is
$a(h_1(\tb_c))=0.90083$.
For $\tb>\tb_c$ the minimum $h_1(\tb)$ becomes negative, and thus
the absolute minimum of the free energy.  This implies that there is a
first-order phase transition at $\tb=\tb_c$, with $\langle \aop\rangle=0$
for $\tb<\tb_c$, and $\langle \aop\rangle\ne 0$ for $\tb>\tb_c$,
with $SU(2)_R$ broken down to $U(1)_R$ in the latter case.

Our mean-field approximation depends, in particular, on the choice of
the mean field.   In order to test the robustness of our conclusions
we performed a
similar analysis, but now starting from a mean field for $\phi$ itself,
rather than for $\phi^\dagger\t_3\phi$.   The details are summarized in
App.~\ref{Alternative}.   Once again we find a first-order transition
to a phase with the same pattern of spontaneous symmetry breaking,
except with a different estimate for the value of $\tb_c$.

\subsection{\label{W mass} The $W$ mass}
In Sec.~\ref{symmetry} we argued that,
much like in the adjoint-Higgs theory \cite{GG,LS}, the spontaneous symmetry
breaking pattern $SU(2)_R\to U(1)_R$ in the reduced model will cause
the $W$ fields to acquire a mass in the full theory.
Using the results from the previous
subsection, we can now check this within the mean-field approximation.
To this end, we restore the external gauge field $e^{igV_{x,\m}}$ of
Eq.~(\ref{Usub}), and expand the action to order $g^2$.   The $\phi$ dependence
is still through the composite operator $\aop(\phi)$ of
Eq.~(\ref{Aphi}), and upon substituting its mean-field value we obtain the
quadratic part of the effective action for the gauge fields in the broken phase.

We find that the terms linear in $V_{x,\m}$ vanish, and those
quadratic in $V_{x,\m}$ are given by
\begin{equation}
\label{Vsq}
\frac{1}{\x}\, a^2\sum_x\left(2d(5a^2-1)\,\tr(W_{x,\mu}^2)+\frac{1}{2}
(a^2+1)\,\tr\left((\partial^-_\m W_{x,\m})^2\right)\right)\ ,
\end{equation}
where $\partial_\m^-$ is the backward derivative.
In this subsection, $a=a(h_{\rm min})$ is the mean-field value
of the order parameter at the global minimum $h_{\rm min}$ of the free energy.
Indeed, a mass term
for the $W$ fields is generated inside the broken phase, where $a\ne 0$.
The $U(1)_R$ gauge invariance implies that,
if higher orders in $g$ would be included, the derivative
$\partial_\m^-$ would change into the derivative $D_\m^-$ covariant
with respect to $U(1)_R$.   Likewise, we observe that no mass term for the
$U(1)_R$ gauge field $A_\m$ is generated.%
\footnote{The role of $U(1)_L$ symmetry is only to turn the field
$\phi$ into an $SU(2)/U(1)$ coset-valued field, as discussed in Sec.~\ref{Higgs}.}

The field $\aop(\phi)$ fluctuates around its mean-field value,
and to go beyond mean-field, this would have to be taken into account.
Some of the fluctuations in $\aop(\phi)$ correspond to the Goldstone
bosons resulting from the symmetry breaking. Other fluctuations,
usually referred to as radial ones, may be generated dynamically.
In general, it is difficult to systematically investigate all possible fluctuations, because the mean-field approximation does not
correspond to the leading order in a systematic expansion in these fluctuations.   However, fluctuations of the field $\aop(\phi)$ that follow
from performing an $SU(2)_R$ gauge transformation on $\phi$ can be
removed by returning to the vector picture, as explained in Sec.~\ref{Higgs}.
In physical terms, this is how the
Goldstone bosons are being ``eaten'' by the massive gauge fields.

Combining Eq.~(\ref{Vsq}) with the transverse kinetic term for the $W$ fields,
one finds the $W$ propagator in momentum space
\begin{equation}
\label{Wprop}
\langle W_{\m,\a}(p)W_{\n,\b}(q)\rangle=
\d(p+q)\d_{\a\b}\left[\left(\d_{\m\n}-\frac{p_\m p_\n}{p^2}\right)
\frac{1}{p^2+m_W^2}+\frac{p_\m p_\n}{p^2}
\frac{\eta}{p^2+\eta m_W^2}\right]\ ,
\end{equation}
with
\begin{equation}
\label{Xi}
\eta=\frac{2\x}{a^2(a^2+1)}\ ,
\end{equation}
and where the $W$ mass is given by
\begin{equation}
\label{Wmass}
m_W^2=2d(5a^2-1)a^2/\x\ .
\end{equation}
The question arises whether or not this is positive in the broken phase.
At the first-order phase transition, the order parameter $a=a(h_{\rm min})$
jumps discontinuously from zero to a different value,
as does the global minimum $h_{\rm min}$ itself.
The new value, which is $a=0.90083$ at the transition,
increases to one for increasing $\tb$.
Thus, $m_W^2$ is positive everywhere in the broken phase.\footnote{%
The same holds in the mean-field approximation considered
in App.~\ref{Alternative}.}

The $W$ mass vanishes in the symmetric phase where $a=0$.
In fact, all of Eq.~(\ref{Vsq}) vanishes in that phase.
In other words, within our mean-field approximation, an effective potential
for the gauge field is generated only in the broken phase, in accordance
with the physical picture advocated in Sec.~\ref{symmetry}.

\subsection{\label{consistency} Discussion of the mean-field solution}
The main weakness of our analysis is that a mean-field
approach does not provide a controlled approximation.  Nevertheless,
we believe that the mean-field calculation is a useful exercise,
since it gives a concrete description of how symmetry breaking might
take place in the equivariantly gauge-fixed theory.
This provides guidance for future
numerical simulations that are the only way to convincingly determine
the phase diagram.

Notwithstanding the
basic limitations of the mean-field solution,
we believe that it passes a number of consistency checks, which are
discussed in the rest of this subsection.

As a preparatory step to the mean-field analysis,
we have computed the effective action~(\ref{Seff})
to leading non-trivial order in $\tb$.
The question arises whether neglecting higher orders is justified.
Put differently, the question is when different orders in the $\tb$-expansion
might start competing with each other.

In order to address this question, we have calculated the $\bc C$ bound-state
propagator within the strong-coupling
expansion in the reduced model by employing the techniques developed
in Ref.~\cite{EP}.  To lowest order in $\tb$ we find for the bound-state propagator
\begin{equation}
\label{CbarCprop}
D(p)=\frac{2}{1+\frac{4\tb}{3}\left(\sum_\m\cos{p_\m}+2\right)}\ .
\end{equation}
The smallest $\tb$ for which the denominator vanishes is
found by setting all the momentum components to $p_\m=\p$.
This happens at $\tb=3/8$.  We take this value as a rough estimate
of the place where successive orders
in the $\tb$-expansion become comparable.    Since
the point $\tb=3/8$ lies deep inside the broken phase we found in mean field,%
\footnote{Both mean-field estimates of the critical value $\tb_c$,
that of Sec.~\ref{reduced} and that of App.~\ref{Alternative},
are much smaller than $3/8$.}
this supports the (self-)consistency
of applying a mean-field approximation to Eq.~(\ref{Seff}).

For technical convenience, we have studied in this section the free energy
$f=f(h,a(h))$ as a function of $h$.  Since the saddle-point
condition~(\ref{mfeqa}) defines a one-to-one mapping between
$h$ and $a$, we may regard the free energy instead as a function of $a$,
$f=f(h(a),a)$.  This function
is interpreted as the mean-field value of the effective potential for
the order parameter $a$.  Thus, even though the reduced model is a topological
field theory, we find that the effective potential has non-trivial
minima.  Given our findings for the toy model of Sec.~\ref{Toy},
this should not come as a surprise.   While the mean-field
estimate of the effective potential may or may not be correct, the topological
nature of the reduced model does not prevent the effective potential from
depending non-trivially on $\tb$.

We have assumed that eBRST symmetry is restored in the thermodynamic limit
of the equivariantly gauge-fixed theory, in both the symmetric and the broken
phases of the reduced model, as well as everywhere in the phase diagram
of the full theory.  Unlike in the toy model, confirming this result
is much more difficult in the field theory case, and goes beyond
the scope of the present paper.
However, it is important to note that the assumption about eBRST symmetry
restoration is not in conflict with anything we know about the broken
phase of the reduced model.  We have already noted in Sec.~\ref{patterns},
that, first, $\vev{\aop}$ is not an order parameter for
eBRST symmetry breaking, and second, that a $W$ mass is compatible
with unbroken eBRST, provided that the ghosts acquire a mass equal to
the $W$ mass.

In the toy model, we have used BRST invariance of the saddle-point expansion
to prove an invariance theorem for $Z_{\pm v}$, from which it follows
that $Z_{\pm v}=1$ to all orders.  This can be interpreted as
the statement that the effective potential for $\phi$ vanishes
at the non-trivial minima $\phi=\pm v$ of the toy model.

Turning to the reduced model,
since we are assuming that eBRST symmetry is not broken spontaneously,
the question arises whether the global minimum of the effective potential
for the order parameter $a$ should always vanish too.  The answer provided
by our mean-field solution is negative.
As explained above, the effective potential is identified
with $f(h(a),a)$, whose global minimum becomes negative in the broken phase.
In order to make sure that
there is no conflict here, it is useful to think in terms
of an effective low-energy lagrangian for the broken phase.
Much like the chiral lagrangian of QCD,
the effective lagrangian $\cl_{\rm eff}$ will be a non-linear
model realizing the symmetry breaking pattern $SU(2)\to U(1)$.
In addition, it should have some type of eBRST symmetry $\hat{s}$
which is inherited from the underlying theory, the reduced model.
Now, in order for the effective theory to satisfy an invariance theorem,
$\cl_{\rm eff}$ would have to be cohomologically exact, namely,
it should have the form of $\hat{s} X$ for some $X$.
In reality, there is no reason why this should be true.
Indeed the reduced-model version of the mass term~(\ref{Smass})
provides an example of a term which is eBRST invariant,
(or, eBRST-closed), and yet it is not eBRST-exact.

\section{\label{phase} Discussion of the phase diagram}
In this section we will discuss what the two-dimensional phase diagram
in the plane spanned by $g$ and $\tg$ may look like.
We rely on what we have learned from the strong-coupling plus mean-field analysis
of the previous sections, augmented by further general considerations.
It should be kept in mind
that there is no guarantee that the mean-field results are correct, but
in this section we will assume that they are.

In the reduced model, \ie, on the $g=0$ boundary, our main result is that
a first-order phase transition occurs going from a symmetric
phase at small $\tb$ to a phase in which the global $SU(2)_R$ symmetry
breaks spontaneously to $U(1)_R$.  We are assuming that eBRST symmetry
is not broken spontaneously in that phase (nor anywhere else in
the phase diagram).  Some considerations supporting this assumption
were presented in Sec.~\ref{patterns} and Sec.~\ref{consistency}.

Since the effective action to which
we applied mean-field techniques was derived in a strong-coupling expansion,
our analysis has nothing to say on what happens near
$\tg=0$ (\ie, $\tb=\infty$).
In particular, we do not know whether or not the broken phase extends
all the way to $\tg=0$.  We will return to this point below.

An important consequence of Eq.~(\ref{Vsq}) is that the first-order phase
transition we found on the $g=0$ boundary extends into the two-dimensional
phase diagram.%
\footnote{
  As explained in Sec.~\ref{symmetry}, for $g>0$ we take the thermodynamic
  limit with an eBRST-breaking seed so as to avoid the invariance theorem.
}
The symmetric phase of the reduced model at small $\tg$ is
the boundary of the familiar confining phase of the full theory.  When, for $g>0$,
the phase-transition line is crossed
towards larger $\tb$ the $W$ fields become massive,
whereas the $A$ field (the ``photon'') stays massless.
At large distances the $W$ fields decouple,
leaving us with an effective abelian theory, which is why we have
referred to this phase as a Coulomb phase.
We conclude that the two phases we found in the reduced model
can be unambiguously differentiated in the full phase diagram as well.
Either the theory is confining with a non-vanishing mass gap;
or there exists a massless photon.

In order to map out possibilities for the full phase diagram, let us consider
the other boundaries, starting with the boundary at $\tg=0$.
Near this boundary it is convenient to rescale $\cw\to g\cw$ in Eq.~(\ref{action}).
The longitudinal kinetic term then has a prefactor $1/\x=g^2/\tg^2$,
which still goes to infinity when $\tg\to 0$ at fixed non-zero $g$.
The four-ghost coupling in Eq.~(\ref{actionb}) goes to zero, while the remaining
terms, which are bilinear in the ghost fields, provide the Faddeev--Popov
determinant appropriate for a maximal abelian gauge (see for example
Ref.~\cite{MAG}).
Near the $\tg=0$ boundary the theory is thus an $SU(2)$ Yang--Mills theory
in a maximal abelian gauge.
We thus expect gauge-invariant observables near the $\tg=0$ boundary
to be the same as in the $SU(2)$ Yang--Mills theory without gauge fixing.
In particular, the theory should be confining,
and possess a mass gap equal to the lowest glueball mass.

Near the $\tb=1/\tg^2=0$ boundary the gauge-fixing sector decouples.
In order to see this, consider the on-shell version of
$S_{\rm gf}$, Eq.~(\ref{onshell}), obtained by integrating over the auxiliary
field $b$.   Rescaling the ghost and anti-ghost fields as
$C\to C/\sqrt{\tg}$, $\bc\to\bc/\sqrt{\tg}$, and taking $\tg\to\infty$,
only the four-ghost term survives.
The gauge-fixing sector decouples, and the theory is again in the confining phase.

Finally, we consider the fourth boundary
of the phase diagram, the one at $\b=0$.   Working to leading order in
both $\tb$ and $\b$, the effective action for $\phi$ is obtained by
integrating over the link variables $U_{x,\m}$ in Eq.~(\ref{Seff}).
For small $\b$ the link variables are randomly distributed,
and the integrals reduce to
simple group integrals.   Carrying out these integrals we find that
$S_{\rm eff}$ reduces to a constant.   This suggests that there is
no phase transition near the $\b=0$ boundary, and that the theory is
again in the confining phase everywhere near this boundary.

This conclusion is supported by the fact that the compact $U(1)$ lattice
theory has a phase transition from a Coulomb phase at weak coupling
to a confining phase at strong coupling.
Below the symmetry-breaking scale originating from the longitudinal
dynamics we have an effective $U(1)$ theory, and thus we should
expect a similar phase transition going towards strong (transversal)
coupling.  Once again, the conclusion is that the Coulomb phase
of the full theory does not extend to the $\b=0$ boundary.

Putting together the information about all four boundaries,
we draw two possible phase diagrams in
Fig.~\ref{phasediag}.  What is common to both panels is the existence
of a single confining phase embracing the Coulomb phase.

\begin{figure}
\includegraphics*[width=7.0cm]{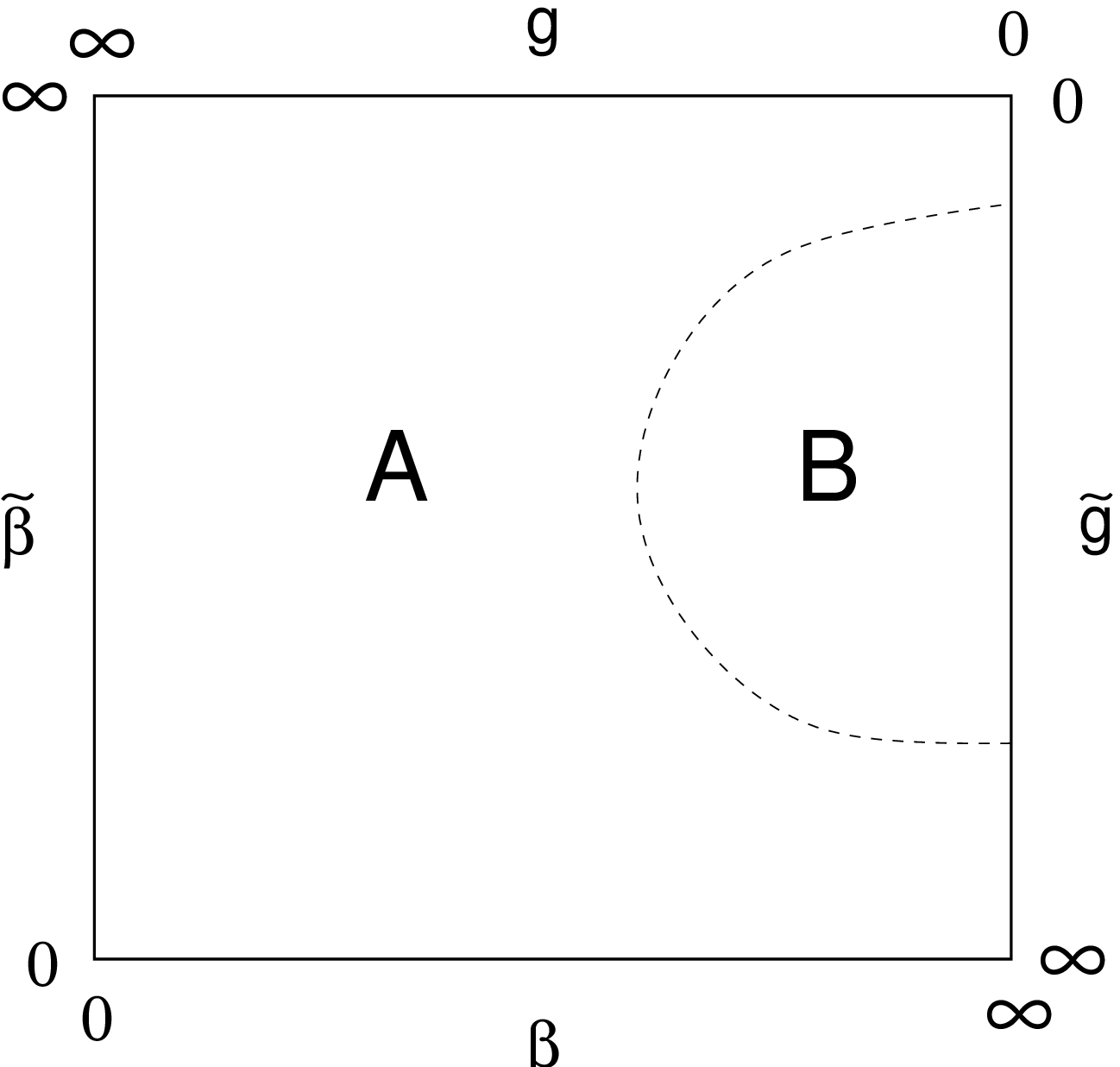}
\hspace{1cm}
\includegraphics*[width=7.0cm]{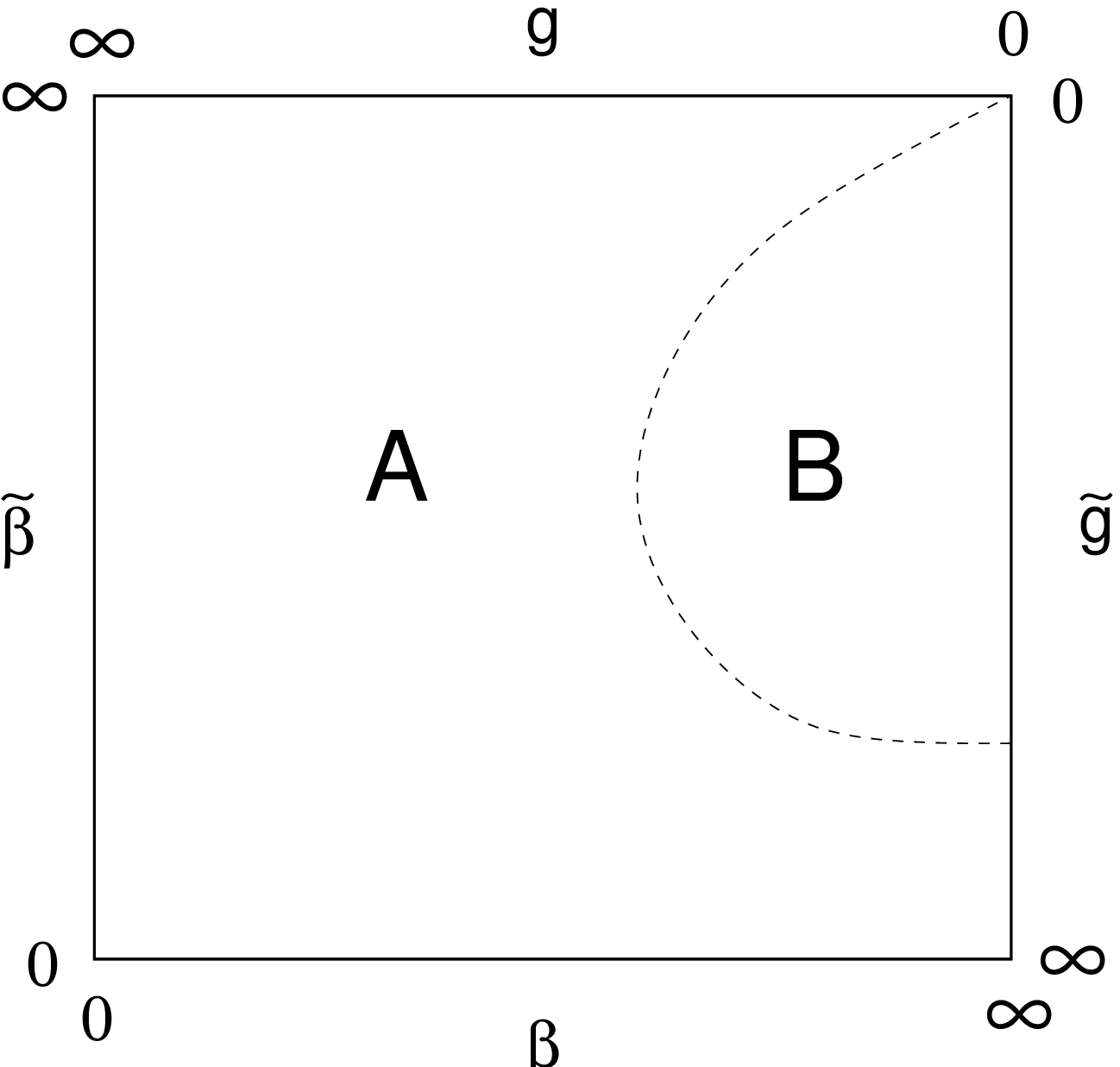}
\vspace*{-2ex}
\begin{quotation}
\floatcaption{phasediag}{%
Two scenarios for the phase diagram.  The confining phase A has a mass gap,
while the Coulomb (or Higgs) phase B has a massless photon.
Left panel: the Coulomb phase ends at some non-zero $\tg$ for $g\to 0$.
Right panel: the Coulomb phase extends to the critical point at $g=\tg=0$.}
\end{quotation}
\vspace*{-3ex}
\end{figure}

The most interesting question concerns the precise structure of the phase
diagram near the gaussian critical point at $(g,\tg)=(0,0)$.
Assuming that the Coulomb phase, predicted by our mean-field approximation,
does indeed exist, the two panels of Fig.~\ref{phasediag}
show the two possibilities.

The panel on the left shows a scenario in which the broken phase
ends at some non-zero $\tg$ for $g\to 0$.   If this is the case, the Coulomb
phase is a lattice artifact.  A more detailed knowledge of its location
and properties is then only important in order to guide numerical simulations
of the theory.

The panel on the right shows a more intriguing scenario,
where both the confining and the Coulomb phases extend to the
asymptotically-free critical point.  The nature of the continuum limit
then depends on how it is taken: from inside the confining phase,
or from inside the Coulomb phase.   The first choice will recover
the standard $SU(2)$ Yang--Mills continuum theory, characterized by
confinement and a mass gap.
By contrast, if the continuum limit is taken from
inside the Coulomb phase, the continuum theory will have massive $W$ gauge
bosons and a massless photon, resembling the broken phase of the
$SU(2)$ theory with a Higgs field in the adjoint representation \cite{GG}.
A novel feature of this scenario is that the Higgs-like behavior
would emerge in a theory in which all couplings are
asymptotically free, and there is no triviality problem.

This speculative scenario raises two main questions.
While answering them is outside the scope of this article,
we offer our current perspective.

The first question is whether there is any
evidence supporting this scenario.   In fact, we believe there is.
In Ref.~\cite{GSb} we derived the one-loop beta function for $\tg$, finding
that it is asymptotically free just like the gauge coupling $g$.  Of course,
the familiar beta function for $g$ does not depend on $\tg$,
whereas the beta function for $\tg$ does depend on both $g$ and $\tg$.
Integrating the renormalization-group equations simultaneously one encounters
dimensional transmutation for both couplings.
We will denote the dynamically generated scales by $\L$ and $\tL$,
respectively.  We choose to define them as the scale where the relevant
coupling becomes equal to one, according to the one-loop running.
In physical terms, $\L$ and $\tL$ are estimates of the energy scales
where the couplings $g$ and $\tg$ become strong.
We remark that the gaussian critical point at $(g,\tg)=(0,0)$
is the unique place in the phase diagram where both of the dynamically
generated scales, $\L$ and $\tL$, tend to zero in lattice units.

The $g$-dependence of the beta function for $\tg$ turns out to have
a non-trivial consequence:  When $g$ becomes strong,
$\tg$ has to become strong, too \cite{GSb}.
In other words, we can have $\tL\gg\L$ or $\tL\sim\L$, but not $\tL\ll\L$.
The exclusion of the latter option leaves us with just two possibilities.
It is thus natural to identify $\tL\sim\L$ with the confining phase,
and $\tL\gg\L$ with the Coulomb phase.

An obvious caveat is that
the notion of the $\L$ parameters is rather elusive, both due
to the freedom in selecting a criterion to define them,
and since we solve the renormalization-group equations in the one-loop approximation.
With this cautionary remark we proceed to discuss the tentative
connection to the phase diagram.

When $\tL\sim\L$, we would not expect the longitudinal dynamics
governed by $\tg$ to qualitatively alter the dynamics of the transverse
degrees of freedom governed by the coupling $g$.
Therefore, if we approach the continuum limit along a trajectory
where the bare $\tg$ is small enough relative to the bare $g$
such that indeed $\tL\sim\L$, we expect to be in the confining phase.
This is consistent with our discussion of the phase diagram near
the boundary $\tg=0$.  Indeed when initially the bare $\tg$ is very small,
its running is primarily driven by that of $g$,
leading to $\tL\sim\L$ \cite{GSb}.

The Coulomb phase would then correspond to choices of bare couplings
such that $\tL\gg\L$.  This hierarchy of scales is natural for the
broken-symmetry phase:  Already at energies large compared to $\L$,
where $g$ is still small, the longitudinal sector becomes strongly interacting
and, presumably, drives the spontaneous symmetry breaking we have found
in mean field.

An important observation is that, while the longitudinal dynamics
that drives the symmetry breaking
must be studied by non-perturbative methods,
the smallness of $g$ allows for a perturbative treatment of
the transverse sector.  In particular, the running of $g$ is still
governed by the one-loop beta function.  However, since the $W$ fields
have acquired a non-zero mass, a mass-independent scheme
would be clearly inappropriate.   A physically sensible definition of a
running coupling would take
the decoupling
of the massive $W$'s into account.   For instance, above the $W$ mass
one can take the running to be defined in a mass-independent scheme
in the full $SU(2)$ gauge theory, while below the $W$ mass it makes more
sense to define the running of the coupling in the surviving effective
abelian theory, with matching of the two couplings at the $W$ mass.

The second question is whether there is any reason to expect that the new
continuum limit taken inside the Coulomb phase would respect the
rules of quantum mechanics and relativity including, in particular, unitarity;
or whether it would just be a curiosity of a statistical system.

At this stage, we can say even less about this second question.
We note that at short distance or high energy unitarity can
be investigated perturbatively, because our theory is asymptotically free
in both couplings.   For a detailed discussion we
refer to Ref.~\cite{gfx}, where we argued that the theory is indeed unitary in
perturbation theory.   However, this does not probe physics in the infrared,
which is where the properties of the two phases are different.

Non-perturbatively, we expect the theory to be unitary in the confining phase,
simply because the physics coincides
with that of unfixed $SU(2)$ Yang--Mills theory, which is unitary.

It is much harder to address the same question in the Coulomb phase.
Within our mean-field approximation, it is
encouraging that both terms in Eq.~(\ref{Vsq}) are positive everywhere in
this phase, leading to a standard massive $W$ propagator, \seef Eq.~(\ref{Wprop}).
According to Refs.~\cite{CLT,LQT} this could therefore lead to a unitary effective
low-energy theory if it is accompanied by a dynamically-generated Higgs field.
What is also likely to be relevant to this question is our expectation
that eBRST symmetry remains unbroken in the Coulomb phase.
This raises the possibility that just as in a ``standard''
Higgs model, also in the Coulomb phase eBRST symmetry provides the tool
to define a projection onto a physical subspace with a unitary S-matrix.

We will postpone further investigations of these questions to the future.
Clearly, first the existence of the Coulomb phase will have to be established more
firmly than is possible with the techniques we employed in this article.

\section{\label{Conclusion} Conclusion and outlook}
In this article, we started an investigation of the phase diagram of equivariantly gauge-fixed $SU(2)$ Yang--Mills theory.   While originally we considered
such theories in the context of a lattice construction of chiral gauge theories,
we believe that they are interesting in their own right,
even without the addition of any fermion fields.
On the one hand, equivariantly gauge-fixed Yang--Mills theories
are well-defined non-perturbatively.  On the other hand,
the transverse and longitudinal gauge couplings, $g$ and $\tg$, are both
asymptotically free, so that at least one of them is expected to become
strong towards the infra-red.  With standard BRST symmetry, these two
couplings are also asymptotically free in common gauges
such as Lorenz gauge or maximal abelian gauge.%
\footnote{
  The $\tg$ beta function for standard BRST gauge fixing in maximal abelian
  gauge \cite{MAG} is the same as for equivariant BRST \cite{GSb}.
}
The key difference is that Yang--Mills theory with standard BRST
gauge-fixing action is not defined outside of perturbation theory \cite{HNnogo}.

Since the goal is to understand this new class of gauge-fixed Yang--Mills theories non-perturbatively,
whether one is ultimately interested in the application to chiral gauge theories
or not, the first order of business is to explore the phase diagram.
Naively, one does not expect the gauge-fixing sector
to alter the physics of the unitary sector of the theory, and therefore one
might expect the whole phase diagram to consist of a single phase---the usual
confining phase.   This expectation is amplified by the
eBRST-based invariance
theorem that was proven in Ref.~\cite{gfx}, extended here in App.~\ref{special},
and reviewed in Sec.~\ref{symmetry}.

However, as we also discussed in Sec.~\ref{symmetry}, there is a loophole.
A richer phase diagram, with potentially physical consequences,
could be uncovered by following a procedure familiar in the study
of spontaneous symmetry breaking:
in order to evade the invariance theorem, a small breaking of eBRST symmetry
is introduced into the finite-volume system,
and is turned off after the thermodynamic limit has been taken.

Our central result concerns the phase diagram of the reduced model.
The latter corresponds to the $g=0$ boundary
of the phase diagram, and inherits from the gauge-fixed Yang--Mills
theory a global symmetry, $SU(2)_R$.  We find that the reduced model
has a phase in which $SU(2)_R$ is broken down to an abelian symmetry.
We discussed circumstantial evidence in support of the assumption that
eBRST symmetry is restored in the limit of vanishing seed.

Furthermore, as we move into the $g>0$ phase diagram,
it is very natural for the broken phase of the reduced model to become
the boundary of a novel phase of the equivariantly gauge-fixed theory:
Two gauge bosons, the $W$'s, ``eat'' the Goldstone bosons arising from
the symmetry breaking, and become massive.  The third
gauge boson, the ``photon,'' stays massless.  Hence the long-distance
physics is that of a Coulomb phase.  We stress that the transverse
sector can be treated perturbatively in $g$, and so it is hard to avoid
this conclusion if indeed the reduced model has
the $SU(2)_R\to U(1)_R$ broken phase.

We have discussed the shortcoming of our analysis.
The main one is that it involves a mean-field
study that does not provide a controlled approximation.  Nevertheless,
the toy model of Sec.~\ref{Toy} teaches us that a topological field theory,
such as the reduced model, can have a non-trivial effective potential,
and thus, a non-trivial phase diagram.  We regard the mean-field calculation
as a means to gain insight into what that phase diagram might be.
Ultimately, the only reliable method for mapping out the
phase diagram is through numerical lattice computations;
the results presented in this paper
provide guidance for initiating a numerical investigation.

Because of the invariance theorem, we know that turning on
an eBRST-breaking seed is a necessary condition for the unveiling
of the phase diagram. This is a novel feature.
In the adjoint-Higgs model, for example, no ``seed'' is needed
in order to probe the Coulomb phase \cite{LS}.  It remains an open
question precisely how the presence of the seed can alter the long-distance
dynamics of the equivariantly gauge-fixed theory.
We observe that the invariance
theorem is, ultimately, a statement about cancellations among
Gribov copies.  Standard arguments show that Gribov copies
can contribute to the partition function with both signs \cite{HNnogo,copies}.
Consequently, the measure of the eBRST gauge-fixed theory can be
both positive and negative.  We conjecture that this fact
is relevant for the dynamical
role of the eBRST-breaking seed as well.  We hope that future
numerical studies will shed light on this question.

If the new Coulomb phase predicted by the mean-field analysis truly exists,
the most conservative view would be to expect it to be a lattice artifact,
disconnected from the continuum limit
defined near the gaussian critical point $(g,\tg)=(0,0)$.   But, based on our
earlier work on the one-loop renormalization-group flow near the critical point \cite{GSb},
we pointed out in Sec.~\ref{phase} that another possibility exists:
the critical point may lie on the boundary separating the two phases.
Which phase is actually realized in the continuum limit would then depend
on how this limit is taken in the $(g,\tg)$ plane,
as we have described in some detail in Sec.~\ref{phase}.

While the possibility that the new phase is connected to the
gaussian critical point is
quite speculative, it is also a very exciting scenario.   If indeed this happens,
and if moreover it could be shown that the continuum limit taken inside the
Coulomb phase is unitary, this would provide us with a novel type of theory
in which all couplings
are asymptotically free, and yet it exhibits the physics of
$SU(2)\to U(1)$ gauge symmetry breaking at low energy.

Clearly, much work remains to be done to establish the existence of a
Coulomb phase, and, then, to investigate its properties.   The results
presented in this article provide us with a framework for setting up a
numerical investigation which we hope to report on in future work.   In addition,
we also plan to investigate whether semi-classical methods can provide us
with more insight by compactifying the theory on one or more spacetime directions and reducing the size of the system in those directions.
Such an approach could be helpful, in particular, in order to find out whether
the Coulomb phase, if it exists, extends to the gaussian fixed point.

We conclude with a few more thoughts on the speculation of a possible
continuum limit with massive $W$'s and a massless photon in the equivariantly
gauge-fixed $SU(2)$ Yang--Mills theory.

First, since in general gauge fixing
is not unique, if such a continuum limit does exist, it appears to open
a Pandora's box of possibilities, making our scenario less attractive
from the point of view of universality.

In fact, our construction of equivariantly gauge-fixed theories
is rather unique.  Consider a (continuum) Yang--Mills theory
based on some gauge group $G$, gauge fixed on the coset
space $G/H$, with $H$ a proper subgroup of $G$.
Since the gauge-fixing action is an integral
part of the non-perturbative definition of the theory,
we demand that it will be Lorentz invariant.
Furthermore, it has to provide a kinetic term for the longitudinal
component of the coset gauge fields, which in turn
should be gauge invariant under the unfixed subgroup $H$.
Together, this implies
that the (on-shell) gauge-fixing action should contain the term
$\tr(D_\m W_\m)^2$, with $D_\m$ the covariant derivative with respect to
the subgroup $H$.   This uniquely fixes the
full gauge-fixing action for the class of theories considered
in App.~\ref{special}, where $G=SU(N)$ and $H$ is a maximal subgroup.
In this case, the non-perturbative construction of the equivariantly
gauge-fixed Yang--Mills theory is thus unique, up to the usual
freedom of changing irrelevant couplings on the lattice.

Finally, we comment on the potential relevance for model building.
At this point, we do not know whether it is possible to extend the
non-perturbative framework
to include the case $G=SU(2)_L\times U(1)_Y$, $H=U(1)_{\rm em}$.
What does fit into the framework of App.~\ref{special} is the choice $G=SU(5)$,
$H=SU(3)\times SU(2)\times U(1)$.  This raises the interesting possibility
that an $SU(5)$ Grand Unified Theory with symmetry breaking down
to $SU(3)_{\rm color}\times SU(2)_L\times U(1)_Y$ might exist without the need
to introduce a Higgs field.

\vspace{3ex}
\noindent {\bf Acknowledgments}
\vspace{3ex}

We thank the referee for raising good questions, and
Jeff Greensite for discussions.
YS thanks the Department of Physics and Astronomy of San Francisco
State University for hospitality.
MG is supported in part by the US Department of Energy, and the Spanish Ministerio de Educaci\'on, Cultura y Deporte, under program SAB2011-0074.
YS is supported by the Israel Science Foundation under grant no.~423/09.
We also thank the Galileo Galilei Institute for Theoretical Physics for
hospitality, and the INFN for partial support.

\appendix
\section{\label{special} Class of equivariantly gauge-fixed Yang--Mills theories}
In Ref.~\cite{gfx} we discussed non-abelian theories with gauge group
$G=SU(N)$, which are equivariantly gauge fixed to a subgroup $H$,
such that the gauge-fixed theory has both eBRST and anti-eBRST symmetry.
In the continuum, the gauge-fixing lagrangian is given by
\begin{equation}
  \cl_{\rm gf} = \ebb s\, \tr \Big(W^2 + \tg^2\, \bc C \Big) \ ,
\label{contL}
\end{equation}
where $W_\m$ is the restriction of the $SU(N)$ vector potential to the coset,
and likewise the ghost fields take values in the coset.
On the lattice, the same definition can be used except that
one has to provide some transcription of the $W^2$ term.
In Ref.~\cite{gfx} this was done for the case that $H$ is the Cartan subgroup.

Here we construct a lattice gauge-fixing action
for all cases where $H$ is a maximal subgroup of $SU(N)$.
A maximal subgroup is uniquely defined by introducing a diagonal matrix
\begin{equation}
  \ttil_0 = \half\; {\rm diag}(\underbrace{1,1,\cdots,1}_{\mbox{$N-M$ times}}, \
                    \underbrace{-1,-1,\cdots,-1}_{\mbox{$M$ times}}\ ) \ .
\label{ttilde}
\end{equation}
The maximal subgroup is the subgroup whose generators commute with $\ttil_0$.%
\footnote{For $N>2$, $\ttil_0$ is a linear combination of the generators
of the Cartan subgroup and of the identity matrix.  The part
proportional to the identity matrix is introduced merely for convenience,
to obtain the suggestive form in Eq.~(\ref{ttilde}).
  }
It is $SU(N-M)\times SU(M) \times U(1)$ for $M>1$,
and $SU(N-1)\times U(1)$ for $M=1$.  The lattice gauge-fixing action is
\begin{equation}
  S_{\rm gf} = \ebb \eb \, \sum_x
  \tr(-2 U_{x,\m} \ttil_0 U_{x,\m}^\dagger \ttil_0 + \tg^2\, \bc_x C_x) \,.
\label{lattgf}
\end{equation}
The eBRST transformation rules retain the simple form~(\ref{ebrst}),
and the anti-eBRST rules again follow via flip symmetry, as discussed
in Sec.~\ref{vector}.
It can be checked that Eq.~(\ref{lattgf}) reduces to Eq.~(\ref{contL})
in the classical continuum limit.  The proof of the invariance theorem,
given in Ref.~\cite{gfx} for the case that $H$ is the Cartan subgroup,
generalizes easily to the case at hand by noting that the Cartan subgroup
is also a subgroup of any maximal subgroup of $SU(N)$.

\section{\label{Toydetails} Proof of Eq.~(\ref{tsaddle2a})}
In order to calculate $Z_0$ we rescale $\phi=\l\psi$, finding
\begin{eqnarray}
Z_0&=&\frac{1}{2\p}\int_{-\infty}^\infty db\int_{-\infty}^\infty d\psi
\,\left(-v^2+3\l^2\psi^2\right)\,e^{-b^2+iv^2b\psi}\,e^{-i\l^2b\psi^3}
\label{tproof}\\
&=&\frac{1}{2\p}\int_{-\infty}^\infty db
\,\sum_{n=0}^\infty\frac{1}{n!}
\,e^{-b^2}\left(\frac{b\l^2}{v^6}\right)^n\left(-v^2-3\,\frac{\l^2}{v^4}\frac{\partial^2}{\partial^2 b}\right)\frac{\partial^{3n}}{\partial^{3n}b}
\int_{-\infty}^\infty d\psi\,e^{iv^2b\psi}
\nonumber\\
&=& \int_{-\infty}^\infty db \,\sum_{n=0}^\infty\frac{1}{n!}
\,e^{-b^2}\left(\frac{b\l^2}{v^6}\right)^n\left(-v^2-3\,\frac{\l^2}{v^4}\frac{\partial^2}{\partial^2 b}\right)\frac{\partial^{3n}}{\partial^{3n}b}\, \d(v^2b)\ .
\nonumber
\end{eqnarray}
In the transition from the first to the second line we have traded
$\psi$ in the interaction term $\exp(-i\l^2b\psi^3)$,
as well as in the ``measure'' term $3\l^2\psi^2$,
with a derivative with respect to $b$.
The integral over $b$ in the last line
is evaluated by integration by parts, using
\begin{equation}
  \frac{\partial^m}{\partial^m b}\left(b^n\,e^{-b^2}\right)\Bigg|_{b=0}
  = (-1)^{\frac{m-n}{2}}\; \frac{m!}{\left(\frac{m-n}{2}\right)!} \ ,
\label{tderivgen}
\end{equation}
which is true for even $m-n\ge 0$ (otherwise the result is zero).  We find
\begin{equation}
\label{tproofcont}
Z_0=-\sum_{n=0}^\infty\frac{(3n)!}{(n!)^2}
\left(\frac{\l^2}{v^6}\right)^n+\sum_{n=0}^\infty\frac{(3(n+1))!}{((n+1)!)^2}
\left(\frac{\l^2}{v^6}\right)^{n+1}=-1\ ,
\end{equation}
where the first and second sums come from the terms with $3n$
and $3n+2$ derivatives respectively.  The final result
comes from the $n=0$ term in the first sum.  For all higher orders
in $\l$ there is a cancellation between the terms in the first and second sum.
This proves Eq.~(\ref{tsaddle2a}) to all orders in $\l$.

\section{\label{Group} Group integrals}
In this Appendix, we collect a few technical details about the calculation
of the integrals in Eqs.~(\ref{uH}) and~(\ref{uH2}).  First, consider Eq.~(\ref{uH}).
Parametrizing the $SU(2)$ matrix
\begin{equation}
\label{phipar}
\phi=x_0+ix_i\t_i\ ,\qquad x^2\equiv x_0^2+x_1^2+x_2^2+x_3^2=1\ ,
\end{equation}
Eq.~(\ref{uH}) takes the form
\begin{equation}
\label{int1}
\frac{1}{\p^2}\int d^4x\,\d(x^2-1)\,e^{h(x_0^2+x_3^2-x_1^2-x_2^2)}\ .
\end{equation}
We introduce new variables
\begin{eqnarray}
\label{newvar}
x_0&=&\sqrt{u}\cos{\c}\ ,\qquad x_3=\sqrt{u}\sin{\c}\ ,\\
x_1&=&\sqrt{v}\cos{\psi}\ ,\qquad x_2=\sqrt{v}\sin{\psi}\ ,\nonumber
\end{eqnarray}
which transforms the integral into
\begin{equation}
\label{int2}
\int_0^1 du\int_0^1 dv\,\d(u+v-1)\,e^{h(u-v)}\ ,
\end{equation}
where the boundaries are a consequence of the delta function.
This integral is easily calculated, and yields the result~(\ref{MFint}).

Next, we also need the integral in Eq.~(\ref{uH2}).
In order to calculate this integral, we first simplify the form of the $2\times 2$
complex matrix $H$.   Using that $H^\dagger H \ge 0$, we can write
\begin{equation}
\label{H2}
H^\dagger H = \L^2\ ,\qquad
\L = R \left(\begin{array}{cc}
           \l_1&0\\
           0&\l_2
       \end{array}\right)
     R^\dagger\ ,
\end{equation}
with real $\l_{1,2}\ge 0$, and $R\in SU(2)$.   Therefore, $H$ can be
written as
\begin{equation}
\label{H}
H=UR \left(\begin{array}{cc}
           \l_1&0\\
           0&\l_2
       \end{array}\right)
     R^\dagger\ ,
\end{equation}
with $U$ unitary.  Since the
Haar measure is both left- and
right-invariant, we can drop $R^\dagger$ on the right and the
$SU(2)$ part of $UR$ on the left, so that
\begin{equation}
\label{Hagain}
H\to e^{i\o} \left(\begin{array}{cc}
           \l_1&0\\
           0&\l_2
       \end{array}\right)
\equiv \left(\begin{array}{cc}
           z_1&0\\
           0&z_2
       \end{array}\right)
\end{equation}
and Eq.~(\ref{uH2}) simplifies to
\begin{equation}
\label{uH22}
e^{u(H)}=\frac{1}{\p^2}\,\int d^4x\,\d(x^2-1)\,e^{(z+z^*)x_0+i(z-z^*)x_3}\ ,
\end{equation}
with $z=(z_1^*+z_2)/2$, and
where we parametrized $\phi$ as in Eq.~(\ref{phipar}).  Using polar
coordinates in four dimensions, with $x_0=r\cos\eta$ and $x_3=r\cos\eta\sin\c$, and performing the integral over all variables except $\eta$, the integral
becomes
\begin{eqnarray}
\label{int0}
e^{u(H)}&=&\frac{1}{i\p y}\int_0^\p d\eta\,\sin\eta\,e^{2x\cos\eta}\,\sin(2iy\sin\eta)\\
&=&\frac{1}{\p x}\int_0^\p d\eta\,\cos\eta\,e^{2x\cos\eta}\,\cos(2iy\sin\eta)\nonumber\\
&=&\frac{1}{4\p x}\frac{\partial}{\partial x}\int_0^{2\p} d\eta\,e^{2x\cos\eta}\,\cosh(2y\sin\eta)\ ,\nonumber
\end{eqnarray}
where we wrote $z=x+iy$.  The second line follows from an integration by
parts, and in the last line we used that the integral over the interval
$[0,\p]$ is equal to that over the interval $[\p,2\p]$.   Finally, switching to
planar polar coordinates for $x$ and $y$ and using periodicity of the
integral over $\eta$, we find that
\begin{equation}
\label{uH2final}
e^{u(H)}=\frac{1}{2h}\frac{\partial}{\partial h}I_0(2h)\left|_{h=\sqrt{x^2+y^2}}=
\frac{1}{h}\,I_1(2h)\right.\ .
\end{equation}
Tracing back, we may identify $h$ with a combination of
invariants of the matrix $H$.   Writing
\begin{equation}
\label{Hpar}
H=\sum_\m (h_\m+ig_\m)\t_\m\ ,\qquad\t_\m=(1,i\t_i)\ ,
\end{equation}
with $h_\m$ and $g_\m$ real,
the combination $x^2+y^2$ in Eq.~(\ref{uH2final}) can be written as
\begin{equation}
\label{repar}
x^2+y^2=\frac{1}{4}\left(\tr(H^\dagger H)+2\,\Re\det\,H\right)=
\sum_\m h_\m^2\equiv h^2\ .
\end{equation}

\section{\label{Alternative} Alternative mean field analysis}
Instead of the composite field of Eq.~(\ref{Aphi}), we may introduce
a mean field for the field $\phi$.
Analogous to Eq.~(\ref{ZMF}), we write
\begin{eqnarray}
\label{Z2}
Z&=&\int \prod_x d_H\phi_x\;e^{-S_{\rm eff}(\phi)}\\
&=&\int \prod_x d_H\phi_x\;e^{-S_{\rm eff}(\phi)}\prod_x\int dV_x\int_{-i\infty}^{i\infty} dH_x\;
e^{-\sum_x\Re\tr(H^\dagger_x(V_x-\phi_x))}\nonumber\\
&=&\prod_x\int dV_x\int dH_x\;e^{-\sum_x\left(\Re\tr(H^\dagger_x V_x)+S_{\rm eff}(V_x)-u(H_x)\right)}\ ,\nonumber
\end{eqnarray}
with
\begin{equation}
\label{uH2}
e^{u(H)}=\int d_H\phi\,e^{\Re\tr(H^\dagger\phi)}\ .
\end{equation}
This integral is calculated in App.~\ref{Group},
and the result is given in Eq.~(\ref{uH2final}).
As in Sec.~\ref{MF}, the mean-field approximation is again obtained
by taking the fields $H$ and $V$ constant,%
\footnote{In this Appendix, $V$ always denotes the constant mean field.}
and evaluating Eq.~(\ref{Z2}) in the saddle-point
approximation, which corresponds to minimizing
the free energy density
\begin{equation}
\label{freeagain}
f(H,V)=\Re\tr(H^\dagger V)+s_{\rm eff}(V)-u(H)\ .
\end{equation}

The action term in $f$ is obtained as follows.
Upon making the replacement~(\ref{Usub}) in Eq.~(\ref{Seff}), one encounters
the combination $\phi_x\phi_x^\dagger$ at several places, for which we will
always substitute the unit matrix.
In all the remaining occurrences we then make the replacement
$\phi\to V$.
We find that $s_{\rm eff}(V)$ is given by Eq.~(\ref{SA}),
where now $A=V^\dagger\t_3 V$.

Using Eq.~(\ref{Hpar}) and, likewise, parametrizing
\begin{equation}
\label{Vpar}
V=\sum_\m(v_\m+iw_\m)\t_\m\ ,
\end{equation}
with $v_\m$ and $w_\m$ real, Eq.~(\ref{freeagain}) leads to the saddle-point
equations
\begin{eqnarray}
\label{spV}
v_\m&=&\frac{1}{2}\frac{\partial u(h)}{\partial h_\m}
=\frac{1}{2}\frac{h_\m}{h}\frac{\partial u(h)}{\partial h}\ ,\qquad w_\m=
\frac{1}{2}\frac{\partial u(h)}{\partial g_\m}=0\ ,\\
h_\m&=&-\frac{1}{2}\frac{\partial s_{\rm eff}(V)}{\partial v_\m}\ ,\qquad g_\m=
-\frac{1}{2}\frac{\partial s_{\rm eff}(V)}{\partial w_\m}\ ,\nonumber
\end{eqnarray}
where we used that $u(H)$ is independent of $g_\m$, \seef Eq.~(\ref{uH2final}).
If we set $w_\m=0$ in Eq.~(\ref{freeagain}), the free energy becomes independent of $g_\m$, and we can thus set $g_\m=0$ as well.

The first equation of Eq.~(\ref{spV}) shows that the direction of $v_\m$ is
the same as that of $h_\m$.
Multiplying both sides of this equation by $\t_\m$, and
writing $h_\m\t_\m=hU_h$ and $v_\m\t_\m=vU_v$ where the $U_{h,v}$ are unitary,
it follows that $U_h=U_v$, and we end up with the simpler equations
\begin{equation}
\label{spsimple}
v=\frac{1}{2}\frac{\partial u(h)}{\partial h}=\frac{I_2(2h)}{I_1(2h)}\ ,
\qquad h=-\frac{1}{2}\frac{\partial s_{\rm eff}(v)}{\partial v}\ .
\end{equation}
We have used that $s_{\rm eff}$ is independent of $U_v$ thanks
of the $SU(2)_R$ invariance of the reduced model.
Note that for $h\to\infty$, $v(h)\to 1$ from below, increasing monotonically
from $v=0$ at $h=0$, reflecting the fact
that the original field $\f_x$ is compact.

\begin{figure}
\begin{center}
\includegraphics*[width=12cm]{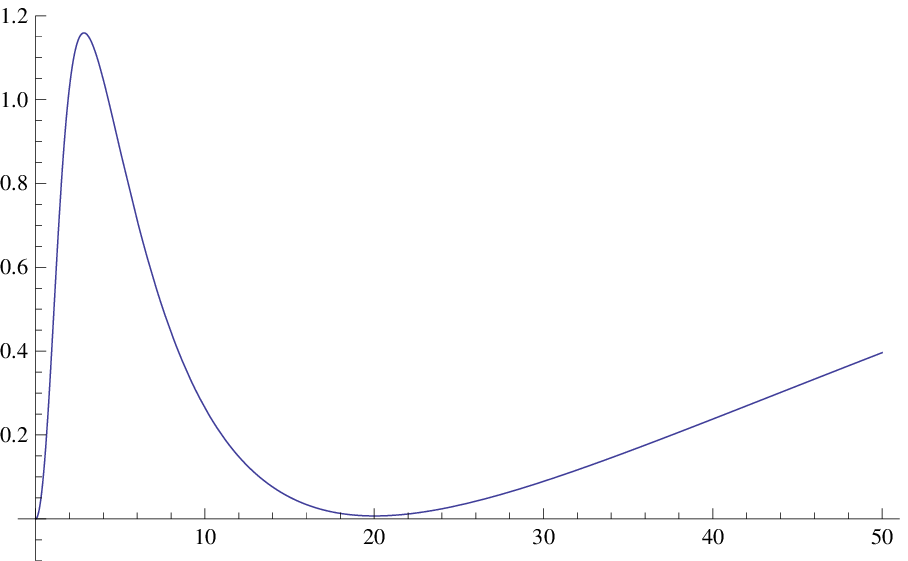}
\end{center}
\begin{quotation}
\floatcaption{free-mfa-new}%
{Free energy~(\ref{freeagain}) as a function of $h$, with $d=4$ and
$\tb=0.09085$.}
\end{quotation}
\vspace*{-4ex}
\end{figure}

We show the free energy $f(h,v(h))$ as a function of $h$ in Fig.~\ref{free-mfa-new}.
Just as in Sec.~\ref{reduced}, there is a first-order transition at a critical
value of the coupling $\tb=\tb_c$, which in this case turns out to be
$\tb_c=0.09085$.   At the transition, $v$ jumps from zero to 0.963,
increasing to one with increasing $\tb$.  Again this leads to a positive
value for $m_W^2$ everywhere in the broken phase.

The method we followed in this Appendix violates Elitzur's theorem \cite{Elitzur} with respect to the $U(1)_L$ gauge group, under which $\phi_x$
is not invariant.   For a detailed discussion of the recovery of gauge
invariance in the presence of a non-zero expectation value for
a gauge non-invariant operator, we refer to Ref.~\cite{DZ}.
In any event, this does not affect the
combination $A=V^\dagger\t_3 V$, which is invariant under $U(1)_L$.
Of course, one can also resort directly to the mean-field treatment we
presented in Sec.~\ref{reduced}.

\section{\label{PT} Perturbative equivalence to standard gauge fixing}
In this Appendix we will prove that an equivariantly gauge-fixed
Yang--Mills theory \cite{gfx} is equivalent to a Yang--Mills theory in a
standard Lorenz gauge at the level of weak-coupling perturbation theory.
With ``equivalent'' we mean that all correlation functions of gauge-invariant
operators are the same between the two theories.   Of course, since
Yang--Mills theory in standard Lorenz gauge is not well defined
outside perturbation theory \cite{HNnogo}, here the equivalence is
necessarily restricted to perturbation theory.   This is
sufficient for our purpose, which is to prove that the mass term of
Eq.~(\ref{Smass}) is not generated in perturbation theory, despite the
fact that it is invariant under (on-shell) eBRST symmetry when $H$ is
a maximal subgroup of $G=SU(N)$.   This
corollary is confirmed by an explicit calculation showing
that no ghost mass is generated at one loop \cite{GZ}.

Using the lattice as a regulator, we start from the action
\begin{equation}
\label{gf1}
S=S_{\rm gauge}(U)+S_{G/H}(U,C,\bc,b)\ ,
\end{equation}
in which $S_{\rm gauge}(U)$ is a lattice discretization of
$\frac{1}{2g^2}\int d^4x\,\tr(F_{\m\n}^2)$, and $S_{G/H}$
is a discretization of the eBRST gauge-fixing action~(\ref{contL}).%
\footnote{For $G=SU(N)$ and $H$ a maximal subgroup, a discretization
that is valid non-perturbatively is given in App.~\ref{special}.
Here, however, we are only interested in perturbation theory, and so
any discretization with the correct classical continuum limit will do.}
In the Higgs picture, the action takes the form
\begin{equation}
\label{gf2}
S=S_{\rm gauge}(U)+S_{G/H}(U^\phi,C,\bc,b)\ ,
\end{equation}
where $U^\phi_{x,\mu}=\phi_xU_{x,\mu}\phi^\dagger_{x+\mu}$ (\seef Eq.~(\ref{gt})).
We recall
that in this picture, $U_{x,\mu}$ is invariant under eBRST transformations,
while $\phi_x$ transforms as given in Eq.~(\ref{ebrstphi}).
The local symmetry of the Higgs picture is $H_L\times G_R$.
The transformation rules are given by Eq.~(\ref{su2r})
generalized to $g_x\in G_R$ and $h_x\in H_L$, where we have
added the the subscripts $R,L$ to indicate from which side the transformation
acts on the $\phi_x$ field.

The Higgs picture has an extra copy $G_R$ of the original local gauge group $G$,
which we will gauge-fix to a standard Lorenz gauge
by adding yet another gauge-fixing term.  Perturbation theory
in the Higgs picture is thus developed from the action
\begin{equation}
\label{gf3}
S=S_{\rm gauge}(U)+S_{G/H}(U^\phi,C,\bc,b)+S_G(U,\z,\bz,\eta)\ ,
\end{equation}
in which the coset-valued field is expanded as
\begin{equation}
\label{phicoset}
  \f_x = \exp\left(i\sqrt{\x}g\, \theta_{x,\a} T_\a \right) \ .
\end{equation}
The restriction of the index $\a$ to the coset generators
eliminates the local $H_L$ invariance.
In Eq.~(\ref{gf3}), $\z$ and $\bz$ are new ghost and anti-ghost fields, and
$\eta$ a new auxiliary field, while
\begin{equation}
\label{var}
S_G = s_G \ck_G(U,\z,\bz,\eta)\ ,
\end{equation}
where $s_G$ is a standard BRST transformation defined for the gauge group $G_R$
in terms of the new ghost and auxiliary fields.
Furthermore, we require that $s$ annihilates $\z$, $\bz$ and
$\eta$, and that $s_G$ annihilates $C$, $\bc$ and $b$.
It follows that $s S_G=0$, because $U$ is invariant under
$s$ in the Higgs picture, \seef Eqs.~(\ref{ebrstphi}) and~(\ref{su2r}).
In addition, $s_G S_{G/H} = 0$, because $S_{G/H}$ depends only
on the combination $U^\phi$.

In Eq.~(\ref{phicoset}) we have reintroduced the gauge-fixing parameter $\x$.
Our next step is to examine the dependence of (un-normalized)
expectation values on this parameter:
\begin{eqnarray}
\label{tstep}
\frac{1}{g^2}\frac{d}{d\x}\langle\co\rangle
&=& \left\langle \co\, s\bs\sum_x\tr\Big(\bc C\Big)\right\rangle\\
&=&-\left\langle (s\co)\, \bs\sum_x\tr\left(\bc C\right)\right\rangle \ ,
\nonumber
\end{eqnarray}
where in the last line we used that $s S_G=0$.
It follows that $\svev{\co}$ is $\x$-independent provided that $s\co=0$,
which is true, in particular, when $\co$ is gauge invariant.

The final step is to observe that, since $\svev{\co}$ does not depend on $\x$,
we may obtain this expectation value order by order in perturbation theory
by considering the $\x\to\infty$ limit.  It is easy to see from Eq.~(\ref{contL})
that, in this limit, the fields $\f$, $C$, $\bc$ and $b$ decouple
from the rest.  The gauge field is controlled by the action
\begin{equation}
\label{gf4}
S=S_{\rm gauge}(U)+S_G(U,\z,\bz,\eta)\ ,
\end{equation}
which is recognized as (lattice discretized) Yang--Mills theory in standard
Lorenz gauge.  The partition function of the decoupled sector containing
the fields $\f$, $C$, $\bc$ and $b$ collapses to a non-zero constant.

We comment that, while the argument based on Eq.~(\ref{tstep}) closely
resembles a key step of the proof of the invariance theorem \cite{gfx},
the discussion in this appendix is restricted to perturbation theory only.
Indeed this is why we may invoke the Lorenz gauge in the first place,
and there is no conflict with the inability to define this gauge
non-perturbatively \cite{HNnogo}.

We conclude that expectation values of gauge-invariant operators
are equal, order by order in perturbation theory,
in the eBRST gauge-fixed theory defined by Eq.~(\ref{gf1}) and in
the Lorenz gauge-fixed
theory defined by Eq.~(\ref{gf4}).
In the latter theory, it is well known that no mass term is generated
perturbatively.  But, had a mass term for (some of) the gluons
been generated in the eBRST gauge-fixed theory~(\ref{gf1}), evidently
this would have altered gauge-invariant correlation functions.%
\footnote{For example, the long-range behavior of
$\svev{F_{\m\n}^2(x)F_{\l\r}^2(y)}$ is sensitive
to the number of massless gluons.}
As a corollary, it follows that the mass term~(\ref{Smass})
cannot be generated perturbatively in the theory defined by Eq.~(\ref{gf1}) either.

\vspace{5ex}

\end{document}